\documentclass[11pt]{article}
\usepackage[final]{acl}

\usepackage{times}
\usepackage{latexsym}
\usepackage[T1]{fontenc}
\usepackage[utf8]{inputenc}
\usepackage{microtype}
\usepackage{inconsolata}
\usepackage{graphicx}

\usepackage{algorithm}
\usepackage{algorithmic}
\usepackage{newfloat}
\usepackage{listings}
\DeclareCaptionStyle{ruled}{labelfont=normalfont,labelsep=colon,strut=off}
\floatstyle{ruled}
\newfloat{listing}{tb}{lst}{}
\floatname{listing}{Listing}
\usepackage{booktabs}
\usepackage{amsmath}
\usepackage{amssymb}
\usepackage{subcaption}
\usepackage{enumitem}
\usepackage{tikz}
\usepackage{colortbl}
\usepackage{etoolbox}
\usepackage[capitalize,noabbrev]{cleveref}

\newcommand{\rev}[1]{#1}
\let\ACLhref\href
\renewcommand{\href}[2]{\ACLhref{#1}{#2}\ (\url{#1})}
\AtBeginEnvironment{thebibliography}{\let\href\ACLhref}
\newcommand{\teaserfigure}{}
\title{AutoMV: Lyric-Grounded Multi-Agent Planning for Full-Song Music Video Generation}
\author{
  \textbf{Xiaoxuan Tang\textsuperscript{2}}
  \thanks{These authors contributed equally to this work.},
  \textbf{Xinping Lei\textsuperscript{2,3,5}}\footnotemark[1],
  \textbf{Chaoran Zhu\textsuperscript{4}}\footnotemark[1],
  \textbf{Shiyun Chen\textsuperscript{1}}
  \\
  \textbf{Ruibin Yuan\textsuperscript{1,6}},
  \textbf{Yizhi Li\textsuperscript{1,7}},
  \textbf{Changjae Oh\textsuperscript{4}},
  \textbf{Ge Zhang\textsuperscript{1}}
  \\
  \textbf{Wenhao Huang\textsuperscript{1}},
  \textbf{Emmanouil Benetos\textsuperscript{4}},
  \textbf{Yang Liu\textsuperscript{2}}
  \thanks{Co-corresponding authors.}
  \\
  \textbf{Jiaheng Liu\textsuperscript{1,3}}\footnotemark[2],
  \textbf{Yinghao Ma\textsuperscript{1,4}}\footnotemark[1]
  \\[0.5em]
  \textsuperscript{1}m-a-p.ai
  \quad
  \textsuperscript{2}Beijing University of Posts and Telecommunications
  \\
  \textsuperscript{3}Nanjing University
  \quad
  \textsuperscript{4}Queen Mary University of London
  \\
  \textsuperscript{5}Zhongguancun Academy
  \\
  \textsuperscript{6}Hong Kong University of Science and Technology
  \\
  \textsuperscript{7}University of Manchester
  \\[0.3em]
  GitHub:
  \ACLhref{https://github.com/multimodal-art-projection/AutoMV}
       {https://github.com/multimodal-art-projection/AutoMV}
  \\
  Website:
  \ACLhref{https://m-a-p.ai/AutoMV/}
       {https://m-a-p.ai/AutoMV/}
}

\begin{document}

\maketitle

\newsavebox{\ACLtabularbox}
\BeforeBeginEnvironment{tabular}{\begingroup\begin{lrbox}{\ACLtabularbox}}
\AfterEndEnvironment{tabular}{%
  \end{lrbox}%
  \ifdim\wd\ACLtabularbox>\linewidth
    \resizebox{\linewidth}{!}{\usebox{\ACLtabularbox}}%
  \else
    \usebox{\ACLtabularbox}%
  \fi
  \endgroup}

\providecommand{\ExpertGroundTruthEqualMean}{3.84}
\providecommand{\ExpertGroundTruthWeightedMean}{3.64}
\providecommand{\ExpertGroundTruthCC}{4.90}
\providecommand{\ExpertGroundTruthPA}{5.00}
\providecommand{\ExpertGroundTruthLS}{4.47}
\providecommand{\ExpertGroundTruthVH}{5.00}
\providecommand{\ExpertGroundTruthSC}{3.73}
\providecommand{\ExpertGroundTruthAC}{3.83}
\providecommand{\ExpertGroundTruthMT}{4.03}
\providecommand{\ExpertGroundTruthST}{3.40}
\providecommand{\ExpertGroundTruthEM}{3.77}
\providecommand{\ExpertGroundTruthVQ}{4.37}
\providecommand{\ExpertGroundTruthCR}{2.48}
\providecommand{\ExpertGroundTruthAN}{1.07}
\providecommand{\ExpertOpenArtEqualMean}{1.78}
\providecommand{\ExpertOpenArtWeightedMean}{1.58}
\providecommand{\ExpertOpenArtCC}{3.60}
\providecommand{\ExpertOpenArtPA}{2.92}
\providecommand{\ExpertOpenArtLS}{1.73}
\providecommand{\ExpertOpenArtVH}{2.97}
\providecommand{\ExpertOpenArtSC}{1.90}
\providecommand{\ExpertOpenArtAC}{1.30}
\providecommand{\ExpertOpenArtMT}{1.18}
\providecommand{\ExpertOpenArtST}{1.15}
\providecommand{\ExpertOpenArtEM}{1.05}
\providecommand{\ExpertOpenArtVQ}{1.38}
\providecommand{\ExpertOpenArtCR}{1.07}
\providecommand{\ExpertOpenArtAN}{1.13}
\providecommand{\ExpertAutoMVEqualMean}{2.26}
\providecommand{\ExpertAutoMVWeightedMean}{2.19}
\providecommand{\ExpertAutoMVCC}{2.58}
\providecommand{\ExpertAutoMVPA}{2.35}
\providecommand{\ExpertAutoMVLS}{2.50}
\providecommand{\ExpertAutoMVVH}{2.92}
\providecommand{\ExpertAutoMVSC}{2.30}
\providecommand{\ExpertAutoMVAC}{2.47}
\providecommand{\ExpertAutoMVMT}{2.68}
\providecommand{\ExpertAutoMVST}{1.83}
\providecommand{\ExpertAutoMVEM}{2.52}
\providecommand{\ExpertAutoMVVQ}{2.47}
\providecommand{\ExpertAutoMVCR}{1.03}
\providecommand{\ExpertAutoMVAN}{1.47}
\providecommand{\ExpertRevidEqualMean}{1.09}
\providecommand{\ExpertRevidWeightedMean}{1.06}
\providecommand{\ExpertRevidCC}{1.00}
\providecommand{\ExpertRevidPA}{2.00}
\providecommand{\ExpertRevidLS}{1.00}
\providecommand{\ExpertRevidVH}{1.00}
\providecommand{\ExpertRevidSC}{1.03}
\providecommand{\ExpertRevidAC}{1.03}
\providecommand{\ExpertRevidMT}{1.03}
\providecommand{\ExpertRevidST}{1.00}
\providecommand{\ExpertRevidEM}{1.00}
\providecommand{\ExpertRevidVQ}{1.00}
\providecommand{\ExpertRevidCR}{1.00}
\providecommand{\ExpertRevidAN}{1.00}
\providecommand{\ExpertAblationFullEqualMean}{2.51}
\providecommand{\ExpertAblationFullWeightedMean}{2.42}
\providecommand{\ExpertAblationFullCC}{3.07}
\providecommand{\ExpertAblationFullPA}{2.95}
\providecommand{\ExpertAblationFullLS}{2.67}
\providecommand{\ExpertAblationFullVH}{3.07}
\providecommand{\ExpertAblationFullSC}{2.00}
\providecommand{\ExpertAblationFullAC}{2.10}
\providecommand{\ExpertAblationFullMT}{3.08}
\providecommand{\ExpertAblationFullST}{2.18}
\providecommand{\ExpertAblationFullEM}{2.60}
\providecommand{\ExpertAblationFullVQ}{3.28}
\providecommand{\ExpertAblationFullCR}{1.23}
\providecommand{\ExpertAblationFullAN}{1.83}
\providecommand{\ExpertAblationLyricsEqualMean}{1.95}
\providecommand{\ExpertAblationLyricsWeightedMean}{1.81}
\providecommand{\ExpertAblationLyricsCC}{2.00}
\providecommand{\ExpertAblationLyricsPA}{2.93}
\providecommand{\ExpertAblationLyricsLS}{2.64}
\providecommand{\ExpertAblationLyricsVH}{3.00}
\providecommand{\ExpertAblationLyricsSC}{2.00}
\providecommand{\ExpertAblationLyricsAC}{1.14}
\providecommand{\ExpertAblationLyricsMT}{2.11}
\providecommand{\ExpertAblationLyricsST}{1.29}
\providecommand{\ExpertAblationLyricsEM}{2.11}
\providecommand{\ExpertAblationLyricsVQ}{2.07}
\providecommand{\ExpertAblationLyricsCR}{1.07}
\providecommand{\ExpertAblationLyricsAN}{1.07}
\providecommand{\ExpertAblationCharacterEqualMean}{1.91}
\providecommand{\ExpertAblationCharacterWeightedMean}{1.84}
\providecommand{\ExpertAblationCharacterCC}{1.22}
\providecommand{\ExpertAblationCharacterPA}{2.37}
\providecommand{\ExpertAblationCharacterLS}{2.70}
\providecommand{\ExpertAblationCharacterVH}{2.85}
\providecommand{\ExpertAblationCharacterSC}{2.00}
\providecommand{\ExpertAblationCharacterAC}{1.15}
\providecommand{\ExpertAblationCharacterMT}{2.54}
\providecommand{\ExpertAblationCharacterST}{1.33}
\providecommand{\ExpertAblationCharacterEM}{2.57}
\providecommand{\ExpertAblationCharacterVQ}{2.04}
\providecommand{\ExpertAblationCharacterCR}{1.11}
\providecommand{\ExpertAblationCharacterAN}{1.07}
\providecommand{\ExpertAblationVerifierEqualMean}{2.10}
\providecommand{\ExpertAblationVerifierWeightedMean}{2.08}
\providecommand{\ExpertAblationVerifierCC}{2.33}
\providecommand{\ExpertAblationVerifierPA}{2.30}
\providecommand{\ExpertAblationVerifierLS}{2.33}
\providecommand{\ExpertAblationVerifierVH}{2.00}
\providecommand{\ExpertAblationVerifierSC}{2.00}
\providecommand{\ExpertAblationVerifierAC}{2.22}
\providecommand{\ExpertAblationVerifierMT}{2.52}
\providecommand{\ExpertAblationVerifierST}{2.11}
\providecommand{\ExpertAblationVerifierEM}{2.54}
\providecommand{\ExpertAblationVerifierVQ}{2.61}
\providecommand{\ExpertAblationVerifierCR}{1.06}
\providecommand{\ExpertAblationVerifierAN}{1.22}

\begin{abstract}
Turning a full song into a music video (MV) requires interpreting its lyrics, which are often metaphorical, multilingual, and repeated across choruses, together with its musical character, and laying the resulting imagery out along the song's structure. Existing music-to-video systems produce short clips or audio-reactive visuals and make little use of lyric meaning. We present AutoMV, a training-free multi-agent framework in which LLM Screenwriter and Director agents plan a full-song visual narrative from time-aligned lyrics and music analysis, keep characters consistent through a shared character bank, and route each shot to cinematic or lip-sync rendering with verification. To evaluate full-song MVs, we build M2VBench: 30 released songs paired with their official MVs, objective audio--visual metrics, and a 12-criterion rubric rated by music-industry experts. Expert evaluation shows that AutoMV outperforms other baselines, although a gap remains to human-produced MVs. Ablation studies highlight the importance of time-aligned lyrics for musical theme alignment and storytelling. We further find that the selected multimodal judge, guided by our evaluation rubric, recovers the expert ranking of automatic systems, with some discrepancies in fine-grained assessments.

\end{abstract}

\begin{figure*}[ht]
  \centering

\includegraphics[width=\linewidth]{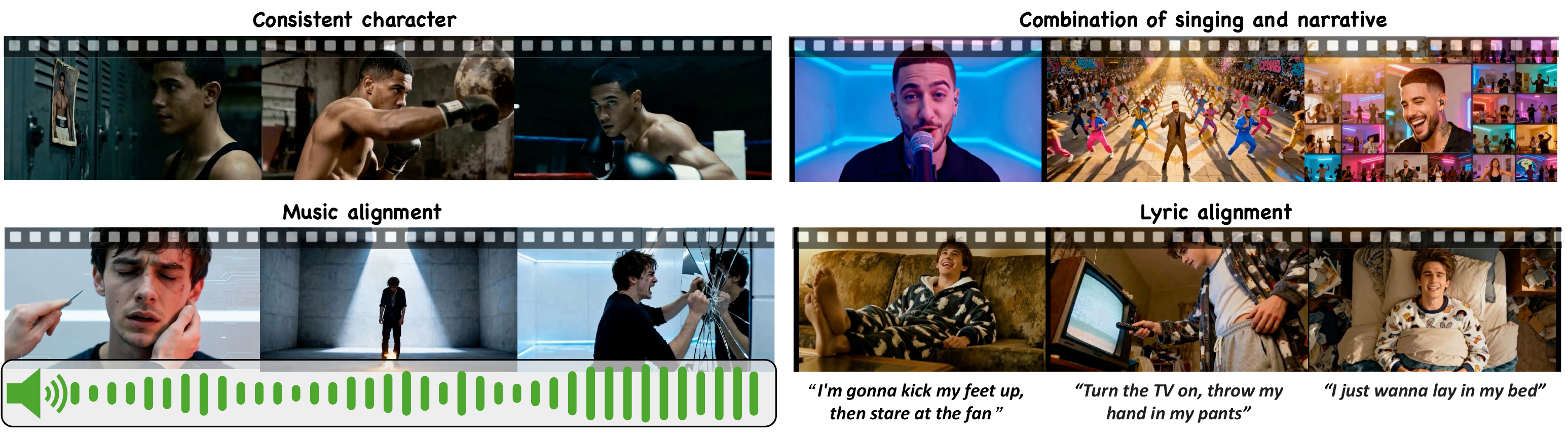}

  \caption{\textbf{AutoMV video generation results.} The framework uses lyric content and musical structure to plan recurring characters, narrative scenes, and singing performances across a full song.}
  \label{fig:teaser}
\end{figure*}

\section{Introduction}
\label{sec:intro}
`
A full-song music video turns musical content into a sequence of visual events. Lyrics, when present, can supply themes, imagery, and narrative perspectives; musical sections and rhythm constrain how those events unfold over time. The resulting video must sustain characters and a narrative across many shots. We study this task as long-form music-to-video (M2V) generation. Advances in image~\citep{gong2025seedream,zhang2025diffusion}, video~\citep{gao2025seedance,henschel2024streamingt2v}, and audio generation~\citep{evans2025stable,qu2024mupt,yuan2025yue} provide its building blocks; connecting them requires planning at the song, section, and shot levels.

Lyric interpretation makes long-form M2V generation a language-understanding problem as well as a generation problem. Lyrics often convey emotion through metaphor rather than literal description, so illustrating each line word for word yields imagery that matches the surface text but misses the song's meaning (\cref{app:failed_metrics}). They also shift narrative perspective, for example between a first-person speaker and an addressed ``you'', which decides who appears on screen and whose story the shots tell. Choruses repeat the same words while their meaning changes with the surrounding verses, and many songs mix languages, such as Chinese and English, within one track, so a planner must read each line in the context of the whole song rather than in isolation.

Professional MV production coordinates scriptwriting, casting, cinematography, and editing at substantial cost~\citep{mv_cost1,mv_cost2}. Automating it requires both song-level interpretation and shot-level temporal coordination. Evaluation poses a related problem: image quality, musical correspondence, and narrative coherence capture different aspects of the result, and multimodal LLM judges are increasingly used to score them, although their agreement with experts on full-length videos is largely untested.

We introduce \textbf{AutoMV}, a training-free multi-agent framework that grounds LLM-based narrative planning in music analysis and time-aligned lyric text. Music information retrieval (MIR) tools produce song structure, captions, separated vocals, and a transcript refined through lyric retrieval. The Screenwriter interprets these inputs to assign themes, actions, and characters to timed scenes. The Director converts the plan into keyframe and video prompts, retrieving shared character descriptions across shots, and routes each shot to cinematic or lip-sync rendering; verification checks generated images and clips before assembly.

We also introduce \textbf{M2VBench}, an evaluation suite of 30 released songs paired with their official MVs. It combines objective metrics for audio--visual correspondence, beat synchronization, and visual quality with a 12-criterion rubric applied by both music-industry experts and multimodal LLM judges. Expert ratings place AutoMV above two commercial systems and below the official MVs.

Our contributions are threefold:
\begin{enumerate}[nosep,leftmargin=*]
    \item \textbf{Lyric-grounded planning for full-song MVs.} AutoMV turns time-aligned lyrics and song structure into a timed script with shared character memory, then renders it through routed cinematic and lip-sync generation with verification.
    \item \textbf{M2VBench with expert annotations.} A full-song benchmark of 30 songs with official MVs as references, objective audio--visual metrics, and 12-criterion ratings from music-industry experts for three automatic systems, three AutoMV ablations, and the official MVs.
    \item \textbf{Empirical findings.} (i) Removing lyric text and timestamps lowers Musical Theme and Storytelling scores, and removing the character bank lowers Character Consistency. (ii) The selected judge ranks automatic systems like experts do but saturates on professional MVs, which limits it as a stand-alone MV evaluator.

\end{enumerate}

\section{Related Work}
\label{sec:related_work}
\subsection{Video generation}
Text-to-video (T2V) models have improved long-form consistency~\citep{henschel2024streamingt2v,bar2024lumiere}, compositional control~\citep{tian2024videotetris}, and photorealism~\citep{lin2024open,gupta2024photorealistic}, supported by large-scale datasets such as OpenVid-1M~\citep{nan2024openvid}. 
While recent models such as 
SeedDance \citep{gao2025seedance}, Google's Veo 3.1\footnote{\href{https://ai.google.dev/gemini-api/docs/video?example=dialogue}{The Google Veo3.1 API document}}, and Wan-2.2 \citep{wan2025wan} generate high-fidelity, multi-shot clips up to 10 seconds with lip-synchronization and audio support, they still lack the temporal consistency and thematic continuity required for full-length storytelling.

To address long-form narrative coherence, recent works explore multi-agent and LLM-guided frameworks. Systems like GenMAC \citep{huang2024genmac}, ViMax\footnote{\href{https://github.com/HKUDS/ViMax/tree/main}{GitHub page - ViMax: Agentic Video Generation}}, 
and HollywoodTown \citep{wei2025hollywood} employ role-based agents (e.g., director, screenwriter) to refine scenes and enforce character consistency across minute-scale videos. Unlike these text-driven approaches, AutoMV aligns visuals to a song's structure, beats, and lyrics.

\subsection{Audio-visual alignment}
Several recent frameworks are proposed for M2V generation. 
MV-Crafter introduces modules for script generation and beat synchronization~\citep{chen2025mv}, CHARCHA focuses on secure and personalized generation~\citep{agarwal2025secure}, and YingVideo-MV employs a music-driven multi-stage approach~\citep{chen2025yingvideo}. 
As these systems are not open-sourced, we benchmark their official demos. The open-source Music2video~\citep{kim2022music2video} focuses on low-level audio reactivity rather than thematic coherence and narrative. 
From Sound to Sight~\citep{vitasovic} provides open-source pipelines for generating music videos from audio-derived scripts, but does not incorporate lyrics and reports limitations in character consistency. AutoMV's lyric transcription and character bank target these limitations. In the reverse direction, LVAS-Agent~\citep{zhang2025long} uses multi-agent collaboration to synthesize audio for long-form videos.

\subsection{Video and Multimodal Evaluation Benchmarks}
\label{subsec:benchmarks}

Visual-centric benchmarks such as VBench~\cite{huang2024vbench}, EvalCrafter~\cite{liu2024evalcrafter}, and Ref4D-VideoBench~\cite{wei2026ref4d} assess spatio-temporal consistency, motion smoothness, and aesthetics, but operate on short, silent clips; even human-directed MVs score poorly on VBench (\cref{app:vbench}).

On the audio side, CLAP~\cite{elizalde2023clap, wu2023large} and MuQ-MuLan~\cite{zhu2025muq} learn joint music--text embeddings, and CMI-RM~\cite{ma2026cmi} scores generated music against heterogeneous instructions, including time-aligned lyrics. These metrics operate only in the audio--text space and cannot assess character consistency, directorial pacing, or visual beat realization.

Joint audio-visual generation is assessed by AVGen-Bench~\cite{zhou2026avgen}, ACVUBench~\cite{yang2025acvubench}, and AGAV-Rater~\cite{cao2025agav}. 
VABench~\citep{hua2026vabench} evaluates joint audio--video generation across semantic correspondence, audio--visual synchronization, and lip--speech consistency.
Omnivideobench has begun addressing the alignment of visuals with long-form audio~\citep{li2025omnivideobench}.
HarmonySet~\citep{zhou2025harmonyset} introduces a video--music dataset and evaluation framework covering rhythmic synchronization, emotional alignment, thematic coherence, and cultural relevance.
M2VBench complements these efforts with full-song generation comparisons, a 12-criterion expert and LLM rubric and song-grouped validation of aggregate judge selection.

\begin{figure*}[ht]
  \centering

\includegraphics[width=\linewidth]{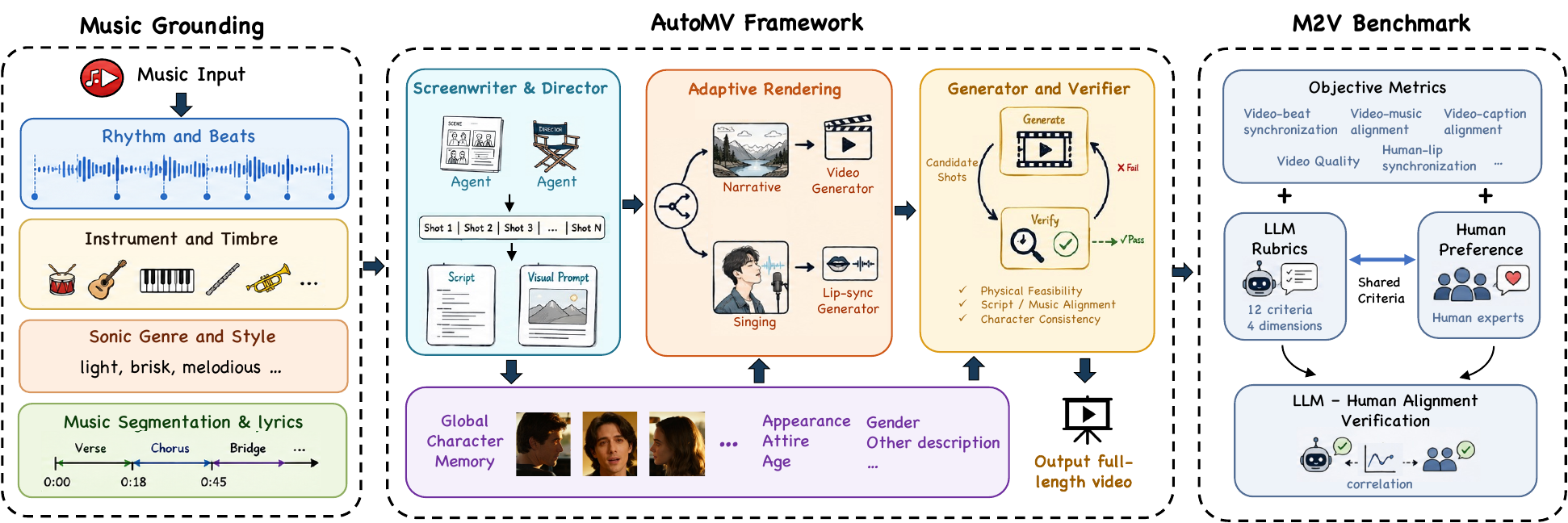}

  \caption{\textbf{Overview.} \textbf{AutoMV} connects music analysis and lyric interpretation to timed scripts, shared character descriptions, and shot-level rendering. Verification checks generated images and clips before assembly. \textbf{M2VBench} evaluates full-song outputs with objective metrics and a 12-criterion rubric, and measures judge agreement with experts using song-grouped validation.}
  \label{fig:pipeline}
\end{figure*}

\section{Methodology}
\subsection{AutoMV workflow}
As illustrated in~\cref{fig:pipeline}, we propose AutoMV, a training-free workflow that transforms full-length songs into coherent, full-length MVs.
It comprises a music-aware goudning system,  a music-to-video generating framework (named AutoMV), and a M2V benchmark, as described below. We refer to each component in the AutoMV by its role throughout, such as Screenwriter and Director; the model instantiating each role is interchangeable and listed in \cref{app:implementation}.

\subsubsection{Step 1: Music and lyric representation}
We represent the song for the planning agents through structured descriptions and timestamps. MIR tools provide four complementary inputs:

    \textbf{Music captioning.} A multimodal model~\citep{xu2025qwen25omnitechnicalreport} summarizes genre, mood, and instrumentation. It also infers vocalist attributes, including gender, which the system uses in character design.
    
    \textbf{Structure analysis.} We use SongFormer~\citep{hao2025songformer} to segment the song structure (intro/verse/chorus/bridge), enabling semantic shot grouping and narrative arcs.
    
    \textbf{Source separation.} With htdemucs~\citep{rouard2023hybrid}, we isolate vocals and accompaniment. The vocal stem supports lyrics transcription and drives lip-sync. 
    
    \textbf{Lyrics transcription.} We utilize Whisper~\citep{radford2023robust} to transcribe the lyrics with timestamps from vocal tracks; the audio track remains the primary input, while a retrieval step refines the transcript and its timestamps when lyrics are available online.

\subsubsection{Step 2: Screenwriting and Directing}
The Screenwriter turns the song representation into timed scenes and shared character descriptions. The Director translates this plan into shot and keyframe prompts. This division connects interpretation of the full song to the local instructions consumed by the rendering models.

\textbf{Temporal layout.} The Screenwriter uses lyric boundaries and musical section transitions to divide the song into contiguous segments. Segment durations are user-controllable, with a default range of 3--15 seconds. Start and end times are quantized to $1/24$\,s to preserve alignment during assembly.

\textbf{Lyric interpretation.} For each timed segment, the Screenwriter interprets the enclosed lyrics in the context of the song. It converts thematic cues, emotional tone, and implied imagery into a description of the scene, atmosphere, and character actions. These descriptions provide the Director with semantic conditions for shot specification and keyframe generation.

\textbf{Character bank}. The Screenwriter instantiates a character label set (detailed appearance, nationality, age, gender) to maintain identity across shots.%
This attribute set is maintained as an external global reference, and both the image generation model and the verifier can perform iterative lookups during the synthesis and validation stages.

\textbf{Shot prompts.} The Director specifies each shot's setting, actions, and camera instructions, then writes a corresponding keyframe prompt. For each character in the prompt, it retrieves the matching bank entry and inserts the shared appearance description.

\textbf{Keyframe images}. The Director requests keyframe image generation from the enriched prompts to seed the video; for subsequent shots, the last frame of the previous clip is reused when continuity is desired. 

\subsubsection{Step 3: Video generation}

The generated keyframes and enriched video prompts divide the long video into subsections. 
At the section level, the Director iteratively verifies keyframes and enriched prompts to enforce consistency before costly video synthesis. At the video level, it routes each section to either the general video renderer or the lip-sync renderer according to the script requirements.

\textbf{
Cinematic rendering}. The video renderer generates 3--8\,s subclips by default. Shots of up to 15\,s combine one to three subclips with aligned frame boundaries. The first subclip uses a Director-generated keyframe; subsequent subclips can use the previous clip's last frame for continuity.

\textbf{Lip-sync rendering}. %
This is used for lines where the Screenwriter judges mouth articulation to be salient
, such as section onsets and chorus entries, 
based on htdemucs vocal stem to the lip-sync backend for video generation, and combine the generated video with the original mixed track.

\subsubsection{Step 4: Verification}
Verification checks text--visual alignment and physical feasibility in two stages. For each keyframe candidate (up to three), the image verifier checks physical plausibility (poses, lighting, and perspective) and adherence to the shot instruction and the character-bank description. For each rendered clip (up to three), the video verifier first rejects visible physical failures, such as broken motion or implausible kinematics. Among the candidates that pass this filter, it scores alignment with the shot action, lyrics, and music caption together with character identity continuity, and accepts the highest-scoring candidate.

\subsection{AutoMV System capabilities and cost}
\begin{table}[h]
  \caption{System capabilities and per-song cost. ``--'' = not disclosed by the system's paper or platform. Multi-char.\ = maintains multiple consistent characters; Lip-sync = lip-synchronized singing; Full-song = generates for the complete song.} 
  \label{tab:capabilities}
  \centering
  \resizebox{\linewidth}{!}{%
  \begin{tabular}{l|cc|ccc}
    \toprule
    Method & Cost(\$) & Time & Multi-char. & Lip-sync & Full-song \\
    \midrule
    \multicolumn{6}{l}{\textit{Commercial Platforms}} \\
    \midrule
    Revid.ai-base & 10 & 5--10 min & $\times$ & $\times$ & \checkmark \\
    OpenArt-story & 20--40 & 10--20 min & $\times$ & \checkmark & \checkmark \\
    \midrule
    \multicolumn{6}{l}{\textit{Academic \& Open-Source Models}} \\
    \midrule
    Music2Video & -- & 5--10 h & $\times$ & $\times$ & \checkmark \\
    CHARCHA & -- & -- & $\times$ & $\times$ & $\times$ \\
    MV-Crafter & -- & -- & $\times$ & $\times$ & $\times$ \\
    AutoMV & 10--20 & 30 min & \checkmark & \checkmark & \checkmark \\
    \midrule
    Human (experts) & $\geq$10k & Weeks & \checkmark & \checkmark & \checkmark \\
    \bottomrule
  \end{tabular}
  }
\end{table}
\cref{tab:capabilities} contrasts AutoMV with the compared systems. AutoMV is the only automatic system that supports all three capabilities required for full-song MV production---multiple consistent characters, lip-synchronized singing, and complete-song generation---at a per-song cost comparable to the cheapest commercial platform.

\subsection{M2VBench}\label{sec:m2vbench}
We instantiate the evaluation of full-song M2V generation on \textbf{M2VBench}. Its song set consists of 30 professionally released songs: 13 in Chinese, 13 in English, and four in Korean, Arabic, or mixed languages; for each song, the full audio track is the system input, and the officially released MV created by professional production teams serves as ground truth. On top of the song set, it comprises three evaluation components, detailed in the next section: (i) objective metrics for visual quality and audio--visual correspondence, (ii) a 12-criterion LLM-judge protocol covering technical, editorial, content, and artistic dimensions, and (iii) an expert human study on the same rubric.

\textbf{Objective evaluation.}
\textbf{ImageBind score (IB)}~\citep{girdhar2023imagebindembeddingspacebind}: cosine similarity between the ImageBind embeddings of the video frames and of the song audio. %
\textbf{CLAP}~\citep{wu2023large}: cosine similarity between the CLAP audio embedding of the song and the text embedding of an MLLM-generated caption of the video. 
\textbf{FVD}~\citep{unterthiner2018towards}: Fr\'echet distance between I3D features of the generated videos (four uniformly spaced 16-frame clips per video) and of the human-directed ground-truth MVs; since the reference is built from our 30 songs, it is reported only for systems generating on the same songs.
\textbf{Beat}: synchronization between musical beats and the optical-flow motion envelope within each lyric-aligned segment. Let $P_s$ be a pulse train formed by Gaussian pulses (standard deviation $0.10$\,s) at the detected beats and let $F_s$ be the optical-flow magnitude sampled at 8\,Hz. For segment $s$, we take the maximum normalized cross-correlation over lags $\ell\in[-6,6]$ samples and average it over valid segments, i.e., $\mathrm{Beat}=|S|^{-1}\sum_{s\in S}\max_{|\ell|\le6}\mathrm{corr}(P_s,F_s[\cdot+\ell])$. We compare Beat scores across systems on the shared 30-song set.
\textbf{LSE-D / LSE-C}~\citep{prajwal2020lip}: SyncNet-based Lip-Sync Error Distance (lower is better) and Confidence (higher is better), which reliably measure lip-sync accuracy in videos; we report them %
when singing characters are included. %
Additional metric diagnostics appear in \cref{app:failed_metrics}.

\textbf{LLM as a judge.}
We use Gemini-3.5-Flash at temperature~0 as the judge. It receives the full-length video together with the song audio and rates it on 12 criteria, defined with music-industry experts, in four categories: Technical, Post-Production, Content, and Art. Because judge scores may not track expert perception, we treat them as secondary to the expert study and test their agreement with experts in \cref{sec:judge_alignment}; five other judge configurations, including the non-Gemini Qwen3.5-Omni, are reported in \cref{app:llm_score}.
Each criterion is scored on a 1--5 scale; category scores average their criteria, and the weighted score combines \textbf{Technical} (20\%: character consistency, physical authenticity, lip sync, visual style harmony), \textbf{Post-Production} (20\%: shot continuity, audio--visual correlation), \textbf{Content} (30\%: musical theme, storytelling, emotional expression), and \textbf{Art} (30\%: visual composition \& quality, creativity, AI novelty), with equal weights within each category.
Full criterion definitions are given in \cref{app:eval}.

\begin{table}[t]
  \caption{\textbf{Objective metrics.} Human-directed MVs provide the reference. FVD and Beat use the shared M2VBench songs (\cref{sec:m2vbench}); CHARCHA and MV-Crafter use external demo songs. LSE requires singing characters. ``--'' = not applicable; \textbf{bold} = best; \underline{underline} = second best. Song-level variability and significance appear in \cref{app:significance}.}
  \label{tab:overall_results}
  \centering
   {%
    \setlength{\tabcolsep}{3pt}
    \renewcommand{\arraystretch}{1.45}
    \resizebox{\linewidth}{!}{%
    \begin{tabular}{l|cccccc}
      \toprule
      Method & IB\,(\%)$\uparrow$ & CLAP\,(\%)$\uparrow$ & FVD$\downarrow$ & Beat %
      \,(\%)$ \uparrow$ & LSE-D$\downarrow$ & LSE-C$\uparrow$ \\
      \midrule
      Human & \underline{24.9} & \textbf{43.3} & ref. & 26.1 & \textbf{8.68} & \textbf{4.62} \\
      \arrayrulecolor{black!40}\midrule\arrayrulecolor{black}
      Music2Video & 11.6 & 12.9 & 2200 & \textbf{28.7} & -- & -- \\
       Revid.ai-base & 20.3 & 34.9 & \underline{1072} & 17.0 & -- & -- \\
       OpenArt-story & 18.8 & 34.7 & 1116 & 24.1 & 9.88 & 3.61 \\
       \textbf{AutoMV} & \textbf{24.9} & \underline{40.9} & \textbf{1050} & \underline{26.7} & \underline{9.79} & \underline{4.53} \\
       \midrule
       \multicolumn{7}{l}{\emph{External demo references (different songs)}} \\
       MV-Crafter & 18.3 & 29.7 & -- & -- & -- & -- \\
       CHARCHA & 7.4 & 31.0 & -- & -- & -- & -- \\
      \bottomrule
    \end{tabular}%
    }
  }
\end{table}

\begin{figure}[t]
  \centering
  {%
\definecolor{radarHuman}{HTML}{E0A526}
\definecolor{radarAuto}{HTML}{1F7A4D}
\definecolor{radarOpen}{HTML}{3B7DD8}
\definecolor{radarRevid}{HTML}{8E6BBF}
\definecolor{radarM2V}{HTML}{C0392B}
\definecolor{radarCharcha}{HTML}{4FA89F}
\definecolor{radarCrafter}{HTML}{C0699B}

\def\Rr{2.05}
\def\radarR#1{{(#1-1)*\Rr/4}}
\newcommand{\radarseries}[3]{%
  \foreach \v [count=\i from 0] in {#3} {\coordinate (#2-\i) at ({90-30*\i}:\radarR{\v});}%
  \draw[#1] (#2-0) \foreach \i in {1,...,11} {-- (#2-\i)} -- cycle;%
}
\newcommand{\lgkey}[1]{\raisebox{0.5ex}{\tikz{\draw[#1] (0,0) -- (0.30,0);}}\hspace{2.2pt}}

\begin{tikzpicture}[line join=round, line cap=round,
                    every node/.append style={text=black},
                    lblfont/.style={font=\sffamily\fontsize{6}{6.8}\selectfont}]
  \foreach \i in {0,...,11}{\draw[black!18, line width=0.25pt] (0,0) -- ({90-30*\i}:\Rr);}
  \foreach \r in {2,3,4}{%
    \draw[black!22, line width=0.25pt, dash pattern=on 1.2pt off 1.2pt] (0,0) circle (\radarR{\r});}
  \draw[black!35, line width=0.4pt] (0,0) circle (\Rr);

  \radarseries{radarM2V,     line width=0.55pt, fill=radarM2V,     fill opacity=0.07}{m2v}{1.10,1.10,1.07,3.43,2.67,3.03,2.50,1.17,2.23,2.43,2.60,3.53}
  \radarseries{radarCrafter, line width=0.55pt, fill=radarCrafter, fill opacity=0.07, dash pattern=on 2pt off 1.3pt}{crafter}{2.91,3.27,1.36,3.82,3.36,4.00,4.27,2.82,3.55,4.27,3.00,3.55}
  \radarseries{radarCharcha, line width=0.55pt, fill=radarCharcha, fill opacity=0.07, dash pattern=on 2pt off 1.3pt}{charcha}{3.24,3.20,2.20,4.16,3.60,4.32,4.20,3.04,3.52,4.20,3.76,4.00}
  \radarseries{radarRevid,   line width=0.55pt, fill=radarRevid,   fill opacity=0.07}{revid}{3.72,3.45,2.12,4.32,3.79,4.32,4.42,3.55,3.92,4.42,3.62,4.06}
  \radarseries{radarOpen,    line width=0.55pt, fill=radarOpen,    fill opacity=0.08}{open}{4.40,3.73,3.43,4.77,4.10,4.43,4.43,3.67,4.00,4.73,3.80,4.13}
  \radarseries{radarHuman,   line width=1.0pt,  fill=radarHuman,   fill opacity=0.13}{human}{4.93,4.90,4.90,4.97,4.97,5.00,4.97,4.47,4.90,4.97,4.57,1.20}
  \radarseries{radarAuto,    line width=1.0pt,  fill=radarAuto,    fill opacity=0.16}{auto}{4.44,4.22,4.45,4.72,4.54,4.91,4.92,4.28,4.53,4.75,4.10,3.90}

  \foreach \lab/\alx/\anc/\dx/\dy [count=\i from 0,
                 evaluate=\i as \ang using {90-30*\i}]
    in {{Character\\Consistency (CC)}/center/south/0/2,
        {Physical\\Authenticity (PA)}/left/{south west}/5/1,
        {Lip Sync (LS)}/left/west/3/2,
        {Visual\\Harmony (VH)}/left/west/0/0,
        {Shot\\Continuity (SC)}/left/west/3/-2,
        {Audio--Visual\\Correlation (AC)}/left/{north west}/5/-1,
        {Musical\\Theme (MT)}/center/north/0/-2,
        {Storytelling (ST)}/right/{north east}/-5/-1,
        {Emotional\\Expression (EM)}/right/east/-3/-2,
        {Visual\\Quality (VQ)}/right/east/0/0,
        {Creativity (CR)}/right/east/-3/2,
        {AI Novelty (AN)}/right/{south east}/-5/1}{%
    \node[anchor=\anc, align=\alx, inner sep=1.4pt, lblfont,
          xshift=\dx pt, yshift=\dy pt] at (\ang:{\Rr+0.14}) {\lab};}

  \node[draw=black!22, rounded corners=2.5pt, line width=0.3pt,
        font=\sffamily\fontsize{6}{7.2}\selectfont, anchor=north, inner sep=4pt]
    at (0,{-\Rr-1.08}) {%
    \begin{tabular}{@{}l@{\hspace{7pt}}l@{\hspace{7pt}}l@{\hspace{7pt}}l@{}}
      \lgkey{radarAuto, line width=1.0pt}\textbf{AutoMV (ours)}
        & \lgkey{radarHuman, line width=1.0pt}Human
        & \lgkey{radarOpen, line width=0.7pt}OpenArt
        & \lgkey{radarRevid, line width=0.7pt}Revid \\[2.2pt]
      \lgkey{radarCharcha, line width=0.7pt, dash pattern=on 2pt off 1.3pt}CHARCHA
        & \lgkey{radarCrafter, line width=0.7pt, dash pattern=on 2pt off 1.3pt}MV-Crafter
        & \lgkey{radarM2V, line width=0.7pt}Music2Video
        & \\
    \end{tabular}};
\end{tikzpicture}%
}
  \caption{\textbf{LLM-judge ratings} from Gemini-3.5-Flash across 12 criteria. Solid lines: systems run on M2VBench; dashed lines: released demo sets. AI Novelty measures purposeful use of AI-native aesthetics. The judge saturates on official MVs and does not separate them from AutoMV, whereas experts do (\cref{sec:judge_alignment}); expert scores in \cref{tab:human_subscores} are the primary comparison. Numerical scores appear in \cref{tab:Gemini-3.5-Flash_subscores}.}
  \label{fig:judge_radar}
\end{figure}

\textbf{Human expert evaluation.}
Experts apply the same 12-criterion rubric.
Three music-industry professionals (record-label practitioners, music-product managers, and MV directors) with at least three years of MV production experience evaluated all videos in the main system comparison and the ablation study on the 12-criterion rubric. Videos were shuffled, presented with anonymized identifiers, and evenly distributed among the raters, each video being rated by one expert (so inter-rater agreement cannot be computed).
Raters watch each full video with its audio and assign 1--5 scores per criterion, aggregated as for the LLM judge into the \emph{expert score}.

\Cref{tab:human_subscores} reports expert scores for all systems and ablations. Song-level comparisons with the LLM judge are described in \cref{app:llm_human_relation}.

\section{Experiments}
\subsection{Experiment setup and baselines}

We run AutoMV, OpenArt-story, Revid.ai-base, and Music2Video on the same 30 full-song audio inputs from M2VBench (\cref{sec:m2vbench}), using each system's generation pipeline. Local inference runs on eight NVIDIA A800 80GB GPUs.

We compare against two commercial, closed-source platforms: OpenArt-story mode\footnote{\href{https://openart.ai/music-video/story}{Narrative video model of OpenArt MV.}} and Revid.ai-base\footnote{\href{https://www.revid.ai/home}{revid.ai}}.
Commercial outputs are generated with manual editing disabled, using modes with comparable per-song costs; as the platforms use their own generation backends, this comparison evaluates complete systems rather than the planning component alone. OpenArt shares AutoMV's image generator (seedream) but uses Hailuo for video, and Revid animates still images into slideshow-style videos without persistent characters or singing.

Music2Video~\citep{kim2022music2video} is reproduced from its released code. We additionally report CHARCHA~\citep{agarwal2025secure} and MV-Crafter~\citep{chen2025mv} as external demo references: their 25 and 11 released videos use the authors' song selections, with median durations of $58$\,s and $35$\,s, respectively. Systems run on M2VBench produce full-song videos with median durations of $244$--$265$\,s. AV-Link and YingVideo-MV release short clips (median $5$\,s and $8$\,s), which fall outside this full-song comparison. The original professionally produced MVs provide the human-directed reference.

\begin{figure*}[htb]
    \centering
    \includegraphics[width=1.0\linewidth]{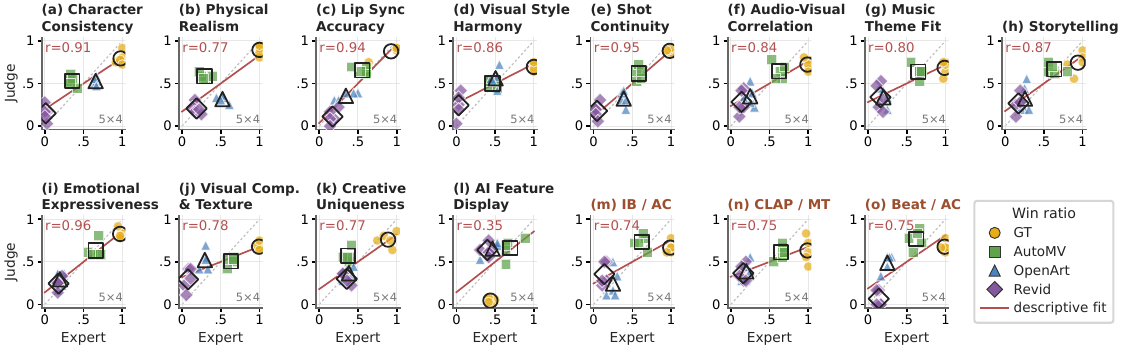}
    \caption{\textbf{Fold-wise method-level human alignment.} Each panel contains 20 fold points: four methods evaluated on five disjoint song folds (six songs per fold). Filled symbols are the individual fold estimates and hollow symbols are the corresponding five-fold means, so the four-method structure is retained while the n-fold replication is visible. The horizontal axis is the expert same-song win ratio and the vertical axis is the corresponding Gemini-3.5-Flash or objective-metric win ratio; wins, ties, and losses contribute $1$, $0.5$, and $0$, respectively. The first twelve panels use the rubric criteria; the final three pair IB and Beat with expert Audio--Visual Correlation (AC), and CLAP with Musical Theme (MT). Red lines and $r$ values are descriptive least-squares fits and Pearson correlations pooled over the 20 fold--method points.}
    \label{fig:alignment35flash}
\end{figure*}

\begin{table*}[ht]
  \caption{
    Detailed expert score across 12 criteria. CC = Character
    Consistency, PA = Physical Authenticity, LS = Lip Sync, VH = Visual Harmony,
    SC = Shot Continuity, AC = Audio--Visual Correlation, MT = Musical Theme,
    ST = Storytelling, EM = Emotional Expression, VQ = Visual Quality,
    CR = Creativity, AN = AI Novelty. The Expert column is the weighted average
    of all criteria (Technical 20\%, Post-Production 20\%, Content 30\%,
    Art 30\%). All scores are in $[1,5]$ (higher is better).
  }
  \label{tab:human_subscores}
  \centering
  \small
    \setlength{\tabcolsep}{8pt}
  \resizebox{\linewidth}{!}{%
  \begin{tabular}{l|cccc|cc|ccc|ccc|c}
    \toprule
    Method &
    CC & PA & LS & VH &
    SC & AC &
    MT & ST & EM &
    VQ & CR & AN &
    Expert
    \\
    \midrule
    \multicolumn{14}{l}{\emph{Primary system comparison.}} \\
    Human (experts) & \ExpertGroundTruthCC & \ExpertGroundTruthPA & \ExpertGroundTruthLS & \ExpertGroundTruthVH & \ExpertGroundTruthSC & \ExpertGroundTruthAC & \ExpertGroundTruthMT & \ExpertGroundTruthST & \ExpertGroundTruthEM & \ExpertGroundTruthVQ & \ExpertGroundTruthCR & \ExpertGroundTruthAN & \ExpertGroundTruthWeightedMean \\
OpenArt-story & \ExpertOpenArtCC & \ExpertOpenArtPA & \ExpertOpenArtLS & \ExpertOpenArtVH & \ExpertOpenArtSC & \ExpertOpenArtAC & \ExpertOpenArtMT & \ExpertOpenArtST & \ExpertOpenArtEM & \ExpertOpenArtVQ & \ExpertOpenArtCR & \ExpertOpenArtAN & \ExpertOpenArtWeightedMean \\
AutoMV & \ExpertAutoMVCC & \ExpertAutoMVPA & \ExpertAutoMVLS & \ExpertAutoMVVH & \ExpertAutoMVSC & \ExpertAutoMVAC & \ExpertAutoMVMT & \ExpertAutoMVST & \ExpertAutoMVEM & \ExpertAutoMVVQ & \ExpertAutoMVCR & \ExpertAutoMVAN & \ExpertAutoMVWeightedMean \\
Revid.ai-base & \ExpertRevidCC & \ExpertRevidPA & \ExpertRevidLS & \ExpertRevidVH & \ExpertRevidSC & \ExpertRevidAC & \ExpertRevidMT & \ExpertRevidST & \ExpertRevidEM & \ExpertRevidVQ & \ExpertRevidCR & \ExpertRevidAN & \ExpertRevidWeightedMean
 \\
    \midrule
    \multicolumn{14}{l}{\emph{Ablation batch; not pooled with the primary system comparison.}} \\
    AutoMV (full; ablation batch) & \ExpertAblationFullCC & \ExpertAblationFullPA & \ExpertAblationFullLS & \ExpertAblationFullVH & \ExpertAblationFullSC & \ExpertAblationFullAC & \ExpertAblationFullMT & \ExpertAblationFullST & \ExpertAblationFullEM & \ExpertAblationFullVQ & \ExpertAblationFullCR & \ExpertAblationFullAN & \ExpertAblationFullWeightedMean \\
AutoMV w/o lyrics & \ExpertAblationLyricsCC & \ExpertAblationLyricsPA & \ExpertAblationLyricsLS & \ExpertAblationLyricsVH & \ExpertAblationLyricsSC & \ExpertAblationLyricsAC & \ExpertAblationLyricsMT & \ExpertAblationLyricsST & \ExpertAblationLyricsEM & \ExpertAblationLyricsVQ & \ExpertAblationLyricsCR & \ExpertAblationLyricsAN & \ExpertAblationLyricsWeightedMean \\
AutoMV w/o character bank & \ExpertAblationCharacterCC & \ExpertAblationCharacterPA & \ExpertAblationCharacterLS & \ExpertAblationCharacterVH & \ExpertAblationCharacterSC & \ExpertAblationCharacterAC & \ExpertAblationCharacterMT & \ExpertAblationCharacterST & \ExpertAblationCharacterEM & \ExpertAblationCharacterVQ & \ExpertAblationCharacterCR & \ExpertAblationCharacterAN & \ExpertAblationCharacterWeightedMean \\
AutoMV w/o verifier & \ExpertAblationVerifierCC & \ExpertAblationVerifierPA & \ExpertAblationVerifierLS & \ExpertAblationVerifierVH & \ExpertAblationVerifierSC & \ExpertAblationVerifierAC & \ExpertAblationVerifierMT & \ExpertAblationVerifierST & \ExpertAblationVerifierEM & \ExpertAblationVerifierVQ & \ExpertAblationVerifierCR & \ExpertAblationVerifierAN & \ExpertAblationVerifierWeightedMean
 \\
    \bottomrule
  \end{tabular}
  }
\end{table*}

\subsection{Main results}

\textbf{Overall performance.} On the shared 30-song set, AutoMV obtains the best automatic-system scores on IB, CLAP, and FVD (\cref{tab:overall_results}); its Beat score is second to Music2Video. Paired comparisons with both commercial baselines show statistically significant improvements in IB, Beat, and the weighted LLM-judge score (\cref{app:significance}).
AutoMV achieves an LSE-C of $4.53$, close to human-directed MVs ($4.62$) and above OpenArt ($3.61$). Music2Video's continuous beat-synchronized transformations yield the highest Beat score ($28.7$); AutoMV ($26.7$) exceeds the human reference ($26.1$), OpenArt ($24.1$), and Revid ($17.0$). Under the Gemini judge (\cref{fig:judge_radar}), AutoMV has the highest score among automatic systems on 10 of the 12 criteria; the judge also scores it close to the official MVs, which experts do not (\cref{sec:judge_alignment}). IB likewise gives AutoMV and human-directed MVs the same mean ($24.9$), whereas the expert comparison separates them (\ExpertAutoMVWeightedMean vs.\ \ExpertGroundTruthWeightedMean); we therefore read IB as a measure of audio--visual correspondence rather than overall quality. 
\textbf{Performance by language.} AutoMV scores above OpenArt and Revid on IB, CLAP, and Beat in both the Chinese and English groups (13 songs each). Its Chinese/English means are $24.71/26.49$ for IB, $39.71/40.01$ for CLAP, and $26.53/25.87$ for Beat. OpenArt's corresponding CLAP means are $39.00/30.36$, showing a larger AutoMV margin on the English songs. \Cref{tab:language_objective_scores} reports means and standard deviations for each method, including the four songs with other or mixed language labels.

\textbf{Human evaluation results.} Experts give AutoMV a weighted score of \ExpertAutoMVWeightedMean, compared with \ExpertOpenArtWeightedMean for OpenArt, \ExpertRevidWeightedMean for Revid, and \ExpertGroundTruthWeightedMean for the official MVs (\cref{tab:human_subscores}). Revid's outputs are static image slideshows without recurring characters, singing, or narrative, so most criteria receive the minimum rating; its row is a lower bound rather than a fine-grained measurement.

\subsection{Do LLM Judges Agree with Experts?}
\label{sec:judge_alignment}
\textbf{Judges saturate on professional MVs.} The Gemini judge gives official MVs and AutoMV the same weighted mean ($4.5$; 95\% CI of the per-song difference $[-0.12, +0.11]$) and awards the maximum rating to 97--100\% of official MVs on five of the twelve criteria (\cref{app:significance}). Expert scores separate the two. The judge therefore compresses the upper end of the quality range, where long-form, lyric-grounded videos differ most from professional work.

\textbf{Judges recover the ranking of automatic systems.} On the weighted score, the Gemini judge orders the expert-rated automatic systems as experts do: AutoMV, then OpenArt, then Revid (\cref{tab:significance}). Because Gemini models also act as AutoMV's Screenwriter and Verifier, we check this ordering with a judge from a different provider: Qwen3.5-Omni ranks AutoMV above OpenArt in all four categories (\cref{tab:Qwen3.5-Omni_subscores}), which reduces the concern that the judge-model overlap favors AutoMV. Among the six cached judge configurations, five are Gemini-family models and one is Qwen3.5-Omni; the selection procedure chooses Gemini-3.5-Flash in all folds. Category-level orderings are less stable; Gemini-3.1-Pro, for example, prefers OpenArt on Technical criteria (\cref{app:llm_score}).

\textbf{Song-level agreement.} \Cref{fig:alignment35flash} repeats the method-level diagnostic over five song-grouped folds; it is descriptive, and the primary human-alignment result is song-level held-out agreement. On the weighted score, the selected judge agrees with the experts' preference on 22 of 30 songs (73.3\%; 95\% CI 55.6--85.8\%). Song-grouped cross-validation over six judge configurations selects Gemini-3.5-Flash in every fold of both the primary five-fold and supplementary three-fold analyses, and the 73.3\% figure is computed on held-out songs, so it does not reflect choosing the judge on the evaluation songs (\cref{app:llm_human_relation}). For reference, fixed objective metrics agree with expert preferences across the three mapped methods (official MVs, OpenArt, AutoMV) at 71.3\% for IB versus Audio--Visual Correlation, 67.2\% for CLAP versus Musical Theme, and 61.5\% for Beat versus Audio--Visual Correlation.

\subsection{Ablation results}

We ablate lyric preprocessing, the character bank, and the verifier; expert scores are in \cref{tab:human_subscores}.

\textbf{Without lyrics \& timestamps.} Removing lyric transcription and timestamp alignment leaves SongFormer structure analysis as the screenwriter's timing input. Within the separate ablation batch, MT falls from $\ExpertAblationFullMT$ to $\ExpertAblationLyricsMT$, ST from $\ExpertAblationFullST$ to $\ExpertAblationLyricsST$, and AC from $\ExpertAblationFullAC$ to $\ExpertAblationLyricsAC$. CC also falls from $\ExpertAblationFullCC$ to $\ExpertAblationLyricsCC$, while PA, LS, and VH remain approximately stable. The changes span content, timing, and character consistency.
\textbf{Without character bank.} Disabling the character bank removes the shared appearance descriptions used across shots. In the ablation batch, CC falls from $\ExpertAblationFullCC$ to $\ExpertAblationCharacterCC$ and PA from $\ExpertAblationFullPA$ to $\ExpertAblationCharacterPA$; experts report mismatched faces, clothing changes, and character swaps. MT and ST also decrease, from $\ExpertAblationFullMT$ to $\ExpertAblationCharacterMT$ and from $\ExpertAblationFullST$ to $\ExpertAblationCharacterST$, respectively. The character bank thus supports both visual consistency and narrative quality.
\textbf{Without verifier.} Without verifier agents, the system accepts more first-round generations. In the ablation batch, Post-Production is largely retained, while PA and VH drop and VQ falls from $\ExpertAblationFullVQ$ to $\ExpertAblationVerifierVQ$. AN also decreases from $\ExpertAblationFullAN$ to $\ExpertAblationVerifierAN$, indicating lower expert ratings for purposeful AI-native aesthetics alongside the decline in visual quality.

\section{Conclusion}
We presented AutoMV, a training-free multi-agent framework that plans full-song music videos from time-aligned lyrics and music analysis with a shared character bank, routed rendering, and verification. On the 30 songs of M2VBench, experts rate AutoMV above two commercial systems and below the official MVs. Removing lyric text and timestamps lowers Musical Theme and Storytelling most among the ablations, and removing the character bank lowers Character Consistency. The selected judge (Gemini-3.5-Flash) reproduces the expert ranking of automatic systems but saturates on professional MVs, so expert evaluation remains necessary. Closing the remaining gap in storytelling, shot continuity, and creativity, and building judges that remain discriminative near professional quality, are the main directions for future work.

\section*{Limitations}
\textbf{Confounded system comparison.} AutoMV and the commercial systems use different generators, so the comparison does not isolate multi-agent planning from backend differences. Isolating it would require feeding single-call LLM scripts to AutoMV's renderers (Doubao and Wan2.2) or running AutoMV's planner on the commercial backends; our ablations instead hold the renderers fixed.

\textbf{Lyric access.} AutoMV may refine its Whisper transcript with retrieved online lyrics, which the commercial systems may not use. The lyric ablation removes transcription and timestamps together, so it does not separate lyric meaning from timing.

\textbf{Evaluation scale and judges.} M2VBench's 30 songs, mostly Chinese and English, limit per-language and per-genre conclusions and make FVD unstable. Gemini models serve as both judge and AutoMV's Screenwriter and Verifier, which could favor AutoMV; the Qwen3.5-Omni results mitigate but do not remove this concern. Songs and MVs may appear in pretraining data, closed APIs can drift, and CHARCHA and MV-Crafter are evaluated on their own, shorter demo songs. In addition, each video is rated by one of three experts, which prevents measuring inter-rater reliability and limits power to detect small differences.

\section*{Ethical Considerations}
The benchmark uses copyrighted songs and official videos. Released materials comprise identifiers or links and derived scores, subject to the applicable licenses and platform terms; commercial audio and video files are excluded. Music-captioning currently infers vocalist attributes that can include gender; this may encode stereotypes and should be replaced by consent-based, task-relevant descriptors. Generated singers can resemble real people, creating portrait, voice, and deepfake risks. We recommend consent checks, provenance labels, and explicit AI disclosure. Lower production costs may change employment patterns for MV professionals; responsible use should preserve human creative direction and consent.

\bibliography{main_acl}

\newpage
\appendix
\raggedbottom
\setcounter{topnumber}{5}
\setcounter{bottomnumber}{3}
\setcounter{totalnumber}{8}
\setcounter{dbltopnumber}{5}
\renewcommand{\topfraction}{0.95}
\renewcommand{\bottomfraction}{0.8}
\renewcommand{\textfraction}{0.05}
\renewcommand{\floatpagefraction}{0.8}
\renewcommand{\dbltopfraction}{0.95}
\renewcommand{\dblfloatpagefraction}{0.8}
\setlength{\floatsep}{12pt plus 2pt minus 2pt}
\setlength{\textfloatsep}{14pt plus 2pt minus 2pt}
\setlength{\dblfloatsep}{12pt plus 2pt minus 2pt}
\setlength{\dbltextfloatsep}{14pt plus 2pt minus 2pt}
\makeatletter
\setlength{\@fptop}{0pt}
\setlength{\@fpsep}{16pt plus 2pt minus 2pt}
\setlength{\@fpbot}{0pt plus 1fil}
\setlength{\@dblfptop}{0pt}
\setlength{\@dblfpsep}{16pt plus 2pt minus 2pt}
\setlength{\@dblfpbot}{0pt plus 1fil}
\makeatother

\providecommand{\rev}[1]{#1}

\begin{figure}
\providecommand{\teaserfigure}{}
\renewcommand{\teaserfigure}{%
  \vskip 0.1in   %
  \begingroup
    \centering
    \includegraphics[width=1.0\linewidth]{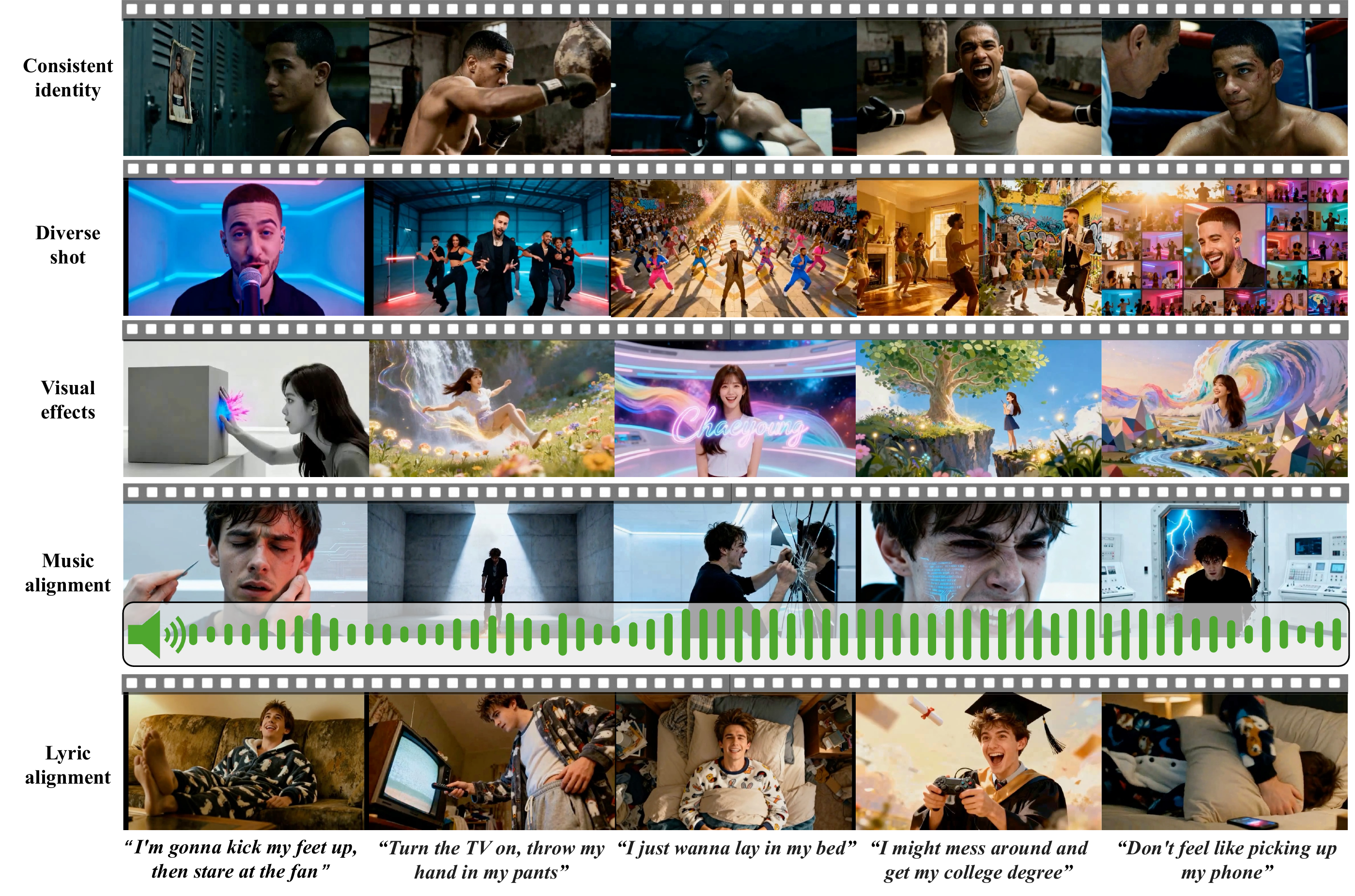}\par
    \captionof{figure}{\textbf{AutoMV video generation results.} We introduce AutoMV, a multi-agent pipeline which produces coherent, music-synchronised full-length music videos guided by beat, structure, and lyric cues. Our pipeline generates music videos which maintain consistent person identity, contain diverse camera shots and visual effects, and align with the corresponding music audio and lyrics.}
    \label{fig:teaser}
  \endgroup
  \vskip 1.5em
}
\end{figure}

\begin{figure}
  \centering

\includegraphics[width=1.0\linewidth]{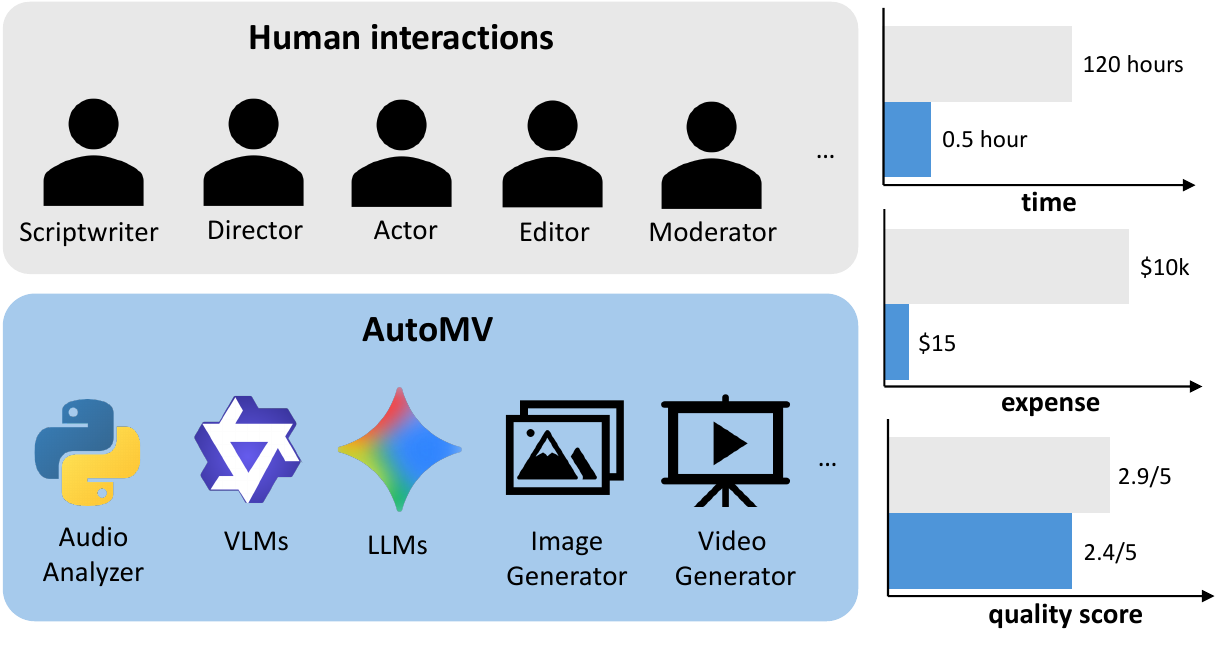}

  \caption{\textbf{Production workflow and reported costs.} AutoMV coordinates music analysis, planning, and generation models. Human-production time and expense estimates are from~\cite{mv_cost1,mv_cost2}; AutoMV uses the reported 30-minute runtime and midpoint of the 10--20 USD cost range. Quality scores are the system-comparison expert means in \cref{tab:human_subscores}.}
  \label{fig:labor}
\end{figure}

\section{Demo Comparison}
The accompanying demos compare full-length outputs from AutoMV, OpenArt-story, and Revid.ai-base. They show the systems' choices of characters, scene content, motion, and lyric-related imagery over the course of a song.

\subsection{Qualitative comparisons}
\begin{figure}[ht]
    \centering
    \includegraphics[width=\linewidth]{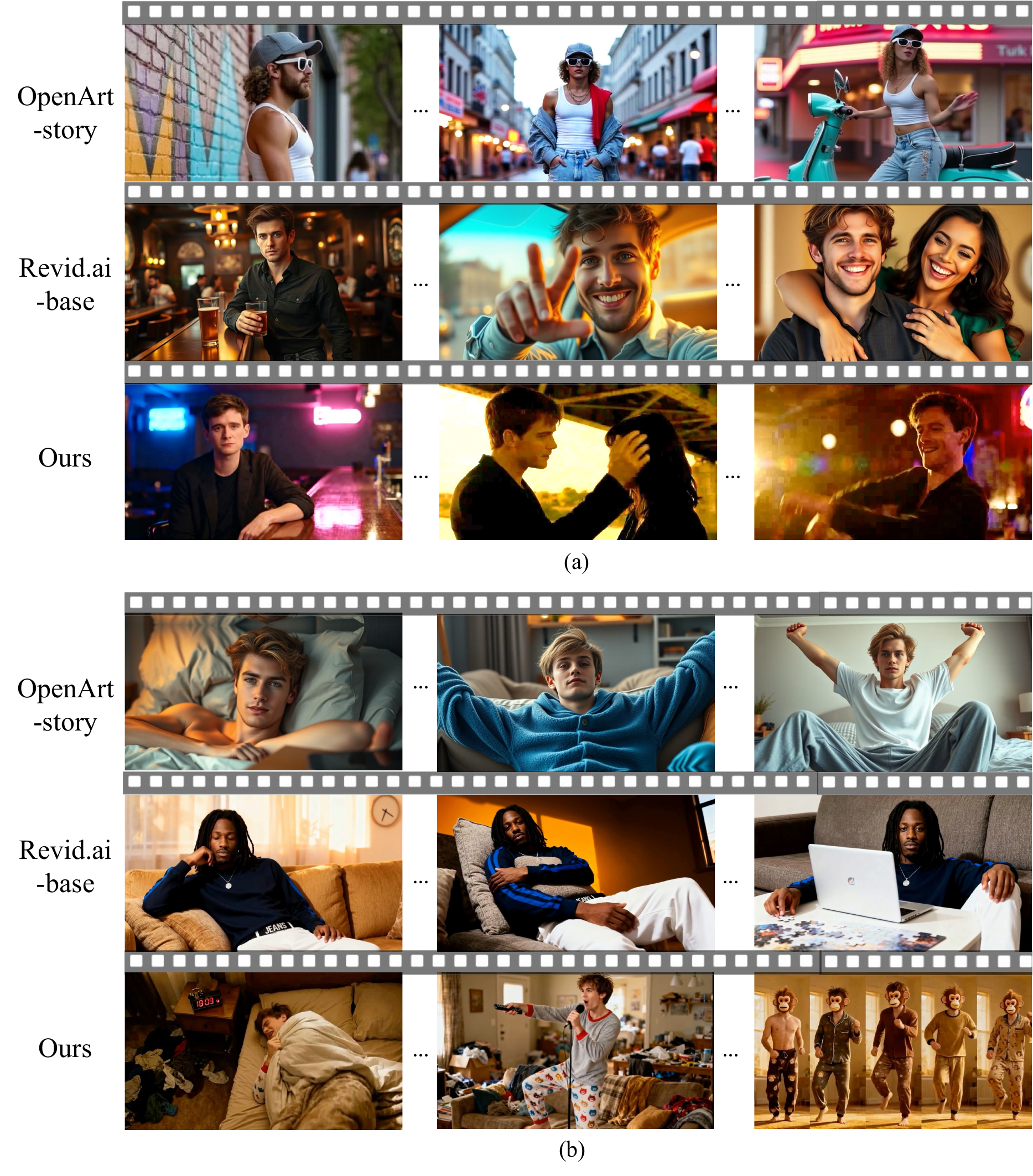}
    \caption{\textbf{Visual comparisons.} Character appearance (a) and scene content (b) in outputs generated from the same input music.}
    \label{fig:comparison}
\end{figure}

\Cref{fig:comparison} shows representative frames from videos generated for the same input music. The AutoMV example in panel (a) retains character appearance and clothing across changes in pose and camera angle. Panel (b) illustrates singing, dancing, and character--environment interactions in the generated scenes.

\section{Detailed Protocol of LLM and Human Experts Scoring}\label{app:eval}
Gemini 2.5 Pro/Flash, Gemini 3 Pro, Gemini 3.1 Pro, Gemini 3.5 Flash, Qwen2.5-Omni, Qwen3-Omni, and Qwen3.5-Omni are prompted with the full-length generated music video and the original audio track. It evaluates 12 sub-criteria, each defined by a five-level rubric describing failure models and ideal behavior. All ratings use a 1–5 integer scale. This rubric is also adopted as the guideline for human evaluation.

\textbf{Four Evaluation Categories}

\textbf{Technical (20\%)}
Average of four sub-points (each weighted 5\% in final score):
\begin{enumerate}
    \item \textbf{Character Consistency:} Stability of appearance across scenes.
    \begin{itemize}
        \item \textbf{1 point:} Character appearance changes frequently, facial features, clothing, and body type show obvious inconsistencies, making it impossible for viewers to confirm it is the same character.
        \item \textbf{2 points:} Character appearance has more than 3 obvious inconsistencies, such as sudden changes in facial features or hairstyle, that significantly affect the viewing experience.
        \item \textbf{3 points:} Character appearance is basically consistent, with 1–2 minor inconsistencies that do not affect the overall viewing experience.
        \item \textbf{4 points:} Character appearance maintains high consistency, and details (such as makeup or accessories) transition naturally between different scenes.
        \item \textbf{5 points:} Character appearance remains perfectly consistent throughout, and even in complex lighting and posture changes, details remain precise.
    \end{itemize}
    
    \item \textbf{Physical Authenticity:} Adherence to real-world physics and natural motion.
    \begin{itemize}
        \item \textbf{1 point:} Serious violation of physical rules throughout, movements are stiff, and objects or characters move against basic physical laws, making it clearly inconsistent with real-world logic.
        \item \textbf{2 points:} Multiple (more than 5) physical inconsistencies are present, such as a floating sensation, model penetration, incorrect object interaction, discontinuous movement, or gravity anomalies.
        \item \textbf{3 points:} Generally follows physical rules, simple movements appear natural, but complex interactions (multi-object collisions, fluids, or fabrics) have obvious flaws.
        \item \textbf{4 points:} Physics effects approach realism, and various object interactions (character movements, environmental elements, or special effects) generally conform to physical rules with only minor details lacking.
        \item \textbf{5 points:} Perfect physical performance, and all element interactions appear as if filmed in reality, including complex movements, environmental physics (water, fire, or smoke), and minute details conforming to natural laws.
    \end{itemize}
    
    \item \textbf{Lip Sync Accuracy:} Alignment between mouth movement and lyrics.
    \begin{itemize}
        \item \textbf{1 point:} Lip movements completely mismatched with lyrics throughout, and they are obviously misaligned or static.
        \item \textbf{2 points:} Over 30\% of lyrical segments have mismatched lip movements, and high notes or special pronunciations have obvious errors.
        \item \textbf{3 points:} Main lyrics are synchronized, but details (such as consonants or prolonged sounds) lack precision, making it approximately 10–20\% asynchronous.
        \item \textbf{4 points:} Over 95\% perfect lip synchronization, including fast-paced segments, with only occasional complex pronunciations showing slight deviations.
        \item \textbf{5 points:} Professional-grade lip synchronization including all syllables, breathing, and emotional changes, which is indistinguishable from a professional singer's live performance.
    \end{itemize}
    
    \item \textbf{Visual Style Harmony:} Color, rendering, and aesthetic consistency.
    \begin{itemize}
        \item \textbf{1 point:} Chaotic visual style, and color tone, texture, and rendering style are completely inconsistent between different scenes.
        \item \textbf{2 points:} Obvious style discontinuities are present, such as realistic scenes suddenly switching to a cartoon style without transition design.
        \item \textbf{3 points:} The style is basically unified, and individual scenes (such as special effects or dream sequences) have stylistic differences but with clear intention.
        \item \textbf{4 points:} The visual style is highly unified throughout, and scene transitions maintain consistent aesthetics and a coordinated color scheme.
        \item \textbf{5 points:} Perfect visual consistency while achieving rich visual layers within a unified style, forming a unique visual identity.
    \end{itemize}
\end{enumerate}
\textbf{Post-Production (20\%)}
Average of two sub-points (each weighted 10\%):
\begin{enumerate}
    \item \textbf{Shot Continuity:} Pacing, transitions, and temporal/spatial coherence.
    \begin{itemize}
        \item \textbf{1 point:} Abrupt shot connections, numerous jump cuts, and chaotic spatial or temporal logic, making it difficult for viewers to follow.
        \item \textbf{2 points:} Lacking basic editing techniques, transitions are crude, rhythm is missing, and there are multiple discontinuous images.
        \item \textbf{3 points:} Uses conventional editing techniques, shot connections are basically smooth, and the rhythm generally matches the music.
        \item \textbf{4 points:} Sophisticated editing techniques, rich shot language, and creative transitions are used, with clear spatial logic and precise rhythm control.
        \item \textbf{5 points:} Master-level editing standard, perfect shot narrative, and every cut has a well-considered artistic purpose, forming a unique editing style.
    \end{itemize}
    
    \item \textbf{Audio–Visual Correlation:} Alignment between visuals and musical rhythm/emotion.
    \begin{itemize}
        \item \textbf{1 point:} Images are almost unrelated to music, and rhythm, mood, and highlight segments are all asynchronous.
        \item \textbf{2 points:} Basic audio-visual synchronization (such as drum beats corresponding to image cuts) is present, but lacking a deeper connection.
        \item \textbf{3 points:} Important musical nodes (chorus or climax) have clear visual correspondence, and basic emotional matching is present.
        \item \textbf{4 points:} Precise audio-visual synchronization, including rhythm, emotional layers, and musical details with visual counterparts.
        \item \textbf{5 points:} Music and images achieve a symbiotic relationship, and visual elements become extensions of the music, forming a unique "audiovisual language."
    \end{itemize}
\end{enumerate}
\textbf{Content (30\%)}
Average of three sub-points (each weighted 10\%):
\begin{enumerate}
    \item \textbf{Musical Theme Relevance:} Alignment of the visuals with the song’s meaning.
    \begin{itemize}
        \item \textbf{1 point:} Video content is completely unrelated to the song theme, and musical emotions contradict visual emotions.
        \item \textbf{2 points:} Superficial relevance (such as literal lyric presentation) is present, but lacking understanding and expression of deeper musical meaning.
        \item \textbf{3 points:} Accurately grasps the core musical theme, and the visual narrative basically matches the song's emotional progression.
        \item \textbf{4 points:} Deeply explores the song's connotations, and enriches musical expression through visual metaphors or symbolic techniques.
        \item \textbf{5 points:} Video becomes an inseparable part of the music, expands musical meaning through a unique perspective, and mutually enhances artistic heights.
    \end{itemize}
    
    \item \textbf{Storytelling:} Narrative structure and character coherence.
    \begin{itemize}
        \item \textbf{1 point:} No clear narrative structure, random image compilation, and viewers are unable to understand the story or theme.
        \item \textbf{2 points:} Has basic narrative elements but confused logic, character motivations are unclear, and plot development is discontinuous.
        \item \textbf{3 points:} Complete narrative structure (beginning-development-climax-conclusion), and basic plot logic is clear.
        \item \textbf{4 points:} Carefully designed narrative framework, character portrayals are well-rounded, plot twists are meaningful, and themes are clear.
        \item \textbf{5 points:} Innovative narrative techniques, multi-layered story structure, and leaves room for thought while maintaining emotional resonance, achieving a short film-level narrative standard.
    \end{itemize}
    
    \item \textbf{Emotional Expression:} Clarity and depth of emotional portrayal.
    \begin{itemize}
        \item \textbf{1 point:} Emotional expression is flat or fake, and character expressions/movements lack emotional conviction, meaning viewers cannot engage.
        \item \textbf{2 points:} Emotional expression is one-dimensional, lacks layered changes, and fails to touch viewers' emotional resonance points.
        \item \textbf{3 points:} Appropriate basic emotional expression, and natural character mood changes can elicit basic empathy from viewers.
        \item \textbf{4 points:} Rich, multi-layered emotional expression, and nuanced emotional changes can trigger deep emotional resonance from viewers.
        \item \textbf{5 points:} Emotional expression achieves artistic sublimation, creates a profound emotional experience through exquisite audiovisual language, and produces a lasting emotional impact on viewers.
    \end{itemize}
\end{enumerate}
\textbf{Art (30\%)}
Average of three sub-points (each weighted 10\%):
\begin{enumerate}
    \item \textbf{Visual Composition \& Quality:} Lighting, framing, and overall fidelity.
    \begin{itemize}
        \item \textbf{1 point:} Image quality is rough, resolution is low, composition is random, and lighting is absent, making the quality similar to early amateur productions.
        \item \textbf{2 points:} Basic image quality is clear, but composition is flat, lighting is simple, and quality is equivalent to an ordinary internet video level.
        \item \textbf{3 points:} Professional standard image quality, standard composition, and basic lighting design, making the quality approaching commercial MV standards.
        \item \textbf{4 points:} High-quality visual presentation, composition and lighting are carefully designed, and color aesthetics are harmonious, resulting in excellent quality.
        \item \textbf{5 points:} Cinematic-level visual quality, every frame is meticulously constructed, and lighting and color achieve artistic standards, allowing it to be appreciated as visual art independently.
    \end{itemize}
    
    \item \textbf{Creativity:} Novelty of concepts, scenes, and transitions.
    \begin{itemize}
        \item \textbf{1 point:} No innovation throughout, and it completely adopts existing MV templates, making the concept highly similar to works from the past three years.
        \item \textbf{2 points:} Only 1–2 common creative points (such as conventional transitions or basic effects) are present, and the core concept lacks uniqueness.
        \item \textbf{3 points:} Thematic innovation is clear (such as narrative structure reorganization), but execution references existing cases.
        \item \textbf{4 points:} Breakthrough visual symbols (such as new camera movement devices) are used, and it achieves conceptual innovation in at least 3 scenes.
        \item \textbf{5 points:} Original worldview throughout, and at least two scenes or camera movements overturn traditional MV design paradigms.
    \end{itemize}
    
    \item \textbf{AI Novelty:} Meaningful integration of AI-native aesthetics.
    \begin{itemize}
        \item \textbf{1 point:} Completely fails to utilize AI technology characteristics, or deliberately imitates traditional filming effects while concealing AI features.
        \item \textbf{2 points:} Passively displays AI characteristics (such as occasional deformation or style breaks), but does not incorporate them into creative design.
        \item \textbf{3 points:} Consciously showcases AI aesthetics (such as style fusion or surreal transformation), but remains at a technical demonstration level.
        \item \textbf{4 points:} Creatively uses AI characteristics as expressive means, forms a unique visual language, and serves narrative or emotional purposes.
        \item \textbf{5 points:} Elevates AI characteristics to artistic language, creates visual "spectacles" impossible with traditional photography while maintaining artistic integrity.
    \end{itemize}
\end{enumerate}

The rubric covers semantic, narrative, and artistic qualities alongside technical criteria. Detailed judge scores appear in \cref{app:llm_score}, and song-level agreement with expert ratings is analyzed in \cref{app:llm_human_relation}.
\section{Implementation Details}\label{app:implementation}

\Cref{tab:implementation} lists the model used for each pipeline role. Generation agents use the providers' public APIs at their default sampling settings; the selected LLM judge runs at temperature~0. The judge configurations compared in the judge-selection analysis are specified in \cref{app:llm_human_relation}. Local music preprocessing and lip-sync rendering run on eight NVIDIA A800 80\,GB GPUs. We estimate generation cost at \$10--\$20.

\begin{table}[!htbp]
  \caption{Model instantiating each role of the AutoMV pipeline.}
  \label{tab:implementation}
  \centering
  \small
  \setlength{\tabcolsep}{4pt}
  \resizebox{\linewidth}{!}{%
  \begin{tabular}{@{}ll@{}}
    \toprule
    Role & Model \\
    \midrule
    Music captioning & Qwen2.5-Omni~\citep{xu2025qwen25omnitechnicalreport} \\
    Structure analysis & SongFormer~\citep{hao2025songformer} \\
    Source separation & htdemucs~\citep{rouard2023hybrid} \\
    Lyrics transcription & Whisper~\citep{radford2023robust} \\
    \midrule
    Screenwriter & Gemini~2.5~Pro~\citep{comanici2025gemini} \\
    Director & \texttt{doubao-1.5-pro}~\citep{seed2025seed1} \\
    Keyframe generation & \texttt{doubao-seedream-4-0-250828}~\citep{seed2025seed1} \\
    General video rendering & Doubao~\citep{seed2025seed1} \\
    Lip-sync rendering & Wan2.2-s2v~\citep{wan2025wan} \\
    Verifiers & Gemini~2.5~Pro~\citep{comanici2025gemini} \\
    \midrule
    LLM judge & Gemini-3.5-Flash, temperature~0 \\
    \bottomrule
  \end{tabular}
  }
\end{table}

\section{Details of Baseline}

\rev{We compare two commercial pipelines, OpenArt-story mode\footnote{\href{https://openart.ai/music-video/story}{OpenArt narrative mode.}} and Revid.ai-base\footnote{\href{https://www.revid.ai/home}{Revid.ai.}}, and reproduce Music2Video from its released code on the benchmark songs. The configurations and external demo references are described below.}

{We use OpenArt's narrative mode for story-driven music videos. Characters are selected from its predefined bank, and the pipeline generates images, renders clips, and composes the final video. Our configuration uses Doubao for image generation and Hailuo for video generation, at around \$20 per song. Doubao is also AutoMV's image generator.}

{Revid.ai-base animates AI-generated images and fits the per-song comparison budget. Revid Pro costs more than three times as much as AutoMV. Revid's internal models are undisclosed.}

For both OpenArt and Revid.ai, we disable all manual editing options to ensure that the generated results are conditioned solely on the input music. Both platforms are run by us on all 30 benchmark songs (30 videos per system), the same protocol as AutoMV and the ground-truth MVs.

{\paragraph{Academic baselines.} We reproduce \emph{Music2Video}~\citep{kim2022music2video} from its released code on all 30 benchmark songs. Its scores, including FVD and Beat, use the same song set as AutoMV. \emph{CHARCHA}~\citep{agarwal2025secure} and \emph{MV-Crafter}~\citep{chen2025mv} contribute external demo references: 25 and 11 released videos on their authors' song selections. Their median durations are $58$\,s (11 of 25 exceed $150$\,s) and $35$\,s (longest $78$\,s), respectively, compared with $244$--$265$\,s for the full-song systems on M2VBench. The table reserves FVD and Beat for the shared benchmark songs. AV-Link and YingVideo-MV release $5$-second audio-to-video snippets and camera-movement ablation excerpts, with median durations of $5$\,s and $8$\,s; these fall outside the full-song comparison. GroundTruth, AutoMV, OpenArt, Revid, and Music2Video each contribute 30 videos.}

\section[Objective Scores by Language]{Objective Scores by Language}\label{app:language_results}
{The language labels define 13 Chinese songs, 13 English songs, and four songs labeled Korean, English/Korean, Arabic/English, or Other. We report equal-weight song means and sample standard deviations for IB, CLAP, and Beat in \cref{tab:language_objective_scores}.}

\begin{table*}[!tp]
\centering
\caption{Objective scores by language group: mean $\pm$ sample standard deviation across songs. All scores use the main table's $100\times$ scale; higher is better. The Other / mixed group contains one Korean, one English/Korean, one Arabic/English, and one Other-labeled song. $^{\dagger}$Music2Video's Chinese CLAP mean uses 12 songs.}
\label{tab:language_objective_scores}
\small
\begin{tabular}{llccc}
\toprule
Language & Method & IB (\%) & CLAP (\%) & Beat (\%) \\
\midrule
Chinese ($n=13$) & Human-directed & $23.47 \pm 4.59$ & $42.93 \pm 13.76$ & $26.20 \pm 6.19$ \\
 & Music2Video & $9.79 \pm 6.22$ & $16.50 \pm 8.28$$^{\dagger}$ & $24.84 \pm 5.29$ \\
 & OpenArt-story & $19.21 \pm 4.95$ & $39.00 \pm 13.04$ & $23.29 \pm 6.14$ \\
 & Revid.ai-base & $20.89 \pm 8.21$ & $38.55 \pm 13.86$ & $15.81 \pm 4.81$ \\
 & AutoMV & $24.71 \pm 6.45$ & $39.71 \pm 11.99$ & $26.53 \pm 3.14$ \\
\midrule
English ($n=13$) & Human-directed & $25.88 \pm 7.38$ & $41.51 \pm 13.30$ & $25.19 \pm 5.05$ \\
 & Music2Video & $12.75 \pm 4.84$ & $10.16 \pm 7.09$ & $29.19 \pm 6.17$ \\
 & OpenArt-story & $18.98 \pm 8.39$ & $30.36 \pm 14.00$ & $23.89 \pm 3.98$ \\
 & Revid.ai-base & $20.55 \pm 5.38$ & $31.50 \pm 10.53$ & $17.24 \pm 5.32$ \\
 & AutoMV & $26.49 \pm 6.32$ & $40.01 \pm 10.69$ & $25.87 \pm 6.89$ \\
\midrule
Other / mixed ($n=4$) & Human-directed & $26.13 \pm 5.46$ & $50.06 \pm 9.04$ & $28.81 \pm 3.76$ \\
 & Music2Video & $13.59 \pm 4.74$ & $10.66 \pm 9.08$ & $39.38 \pm 5.72$ \\
 & OpenArt-story & $17.20 \pm 6.42$ & $34.62 \pm 19.75$ & $27.39 \pm 5.39$ \\
 & Revid.ai-base & $17.44 \pm 8.02$ & $33.93 \pm 16.31$ & $19.86 \pm 4.95$ \\
 & AutoMV & $20.40 \pm 5.99$ & $47.91 \pm 7.67$ & $29.65 \pm 5.40$ \\
\bottomrule
\end{tabular}
\end{table*}

\section{Detailed LLM-score Results by Category}\label{app:llm_score}

{Tables~\ref{tab:Gemini-2.5-Flash_subscores}, \ref{tab:Gemini-2.5-Pro_subscores}, \ref{tab:Gemini-3-pro-preview_subscores}, \ref{tab:Qwen-Omni-3_subscores}, \ref{tab:Qwen-Omni-2.5_subscores}, \ref{tab:Gemini-3.1-Pro_subscores}, \ref{tab:Gemini-3.5-Flash_subscores}, and~\ref{tab:Qwen3.5-Omni_subscores} report the eight judges' criterion scores. Qwen-Omni-3 and Qwen-Omni-2.5 process a 30-second clip from the middle of each video, covering approximately 4--6 shots and consuming around 18k tokens with audio and prompts under a 20k-token input limit. Gemini-3.1-Pro, Gemini-3.5-Flash, and Qwen3.5-Omni process full-length videos through gateways, compressed to 480p at 8\,fps with audio retained.}
\begin{table*}[!tp]
  \caption{
    Detailed Gemini-2.5-Flash evaluation across 12 sub-criteria and 4 category scores.
    Abbr. of 12 {sub-criteria}: CC = Character Consistency, PA = Physical Authenticity,
    LS = Lip Sync, VH = Visual Harmony;
    SC = Shot Continuity, AC = Audio--Visual Correlation;
    MT = Musical Theme, ST = Storytelling, EM = Emotional Expression; 
    VQ = Visual Quality, CR = Creativity, and AN = AI Novelty.
    Abbr. of {categories}: Tech = Technical, Post = Post-Production, and Cont = Content.
    All scores are in $[1,5]$ (higher is better).
  }
  \label{tab:Gemini-2.5-Flash_subscores}
  \centering
  \setlength{\tabcolsep}{3pt}
  \resizebox{\linewidth}{!}{%
  \begin{tabular}{l|cccc|cc|ccc|ccc|cccc}
    \toprule
    Method &
    CC & PA & LS & VH &
    SC & AC &
    MT & ST & EM &
    VQ & CR & AN &
    Tech & Post & Cont & Art \\
    \midrule
    OpenArt-story &
    4.77 & 3.93 & 3.00 & 4.63 &
    4.07 & 4.43 &
    4.60 & 3.43 & 4.10 &
    4.63 & 3.90 & 4.10 &
    4.19 & 4.25 & 4.04 & 4.21 \\
    AutoMV (full) &
    4.13 & 4.33 & 4.13 & 4.43 &
    4.10 & 4.83 &
    4.87 & 4.00 & 4.80 &
    4.73 & 3.77 & 2.67 &
    4.26 & 4.47 & 4.56 & 3.72 \\
    \midrule
    Human (experts) &
    4.87 & 4.53 & 4.37 & 4.53 &
    4.07 & 4.90 &
    5.00 & 3.97 & 4.67 &
    4.50 & 4.03 & 1.27 &
    4.58 & 4.49 & 4.55 & 3.27 \\
    \bottomrule
  \end{tabular}
    }
\end{table*}

\begin{table*}[!tp]
  \caption{
    Detailed Gemini-2.5-Pro evaluation across 12 sub-criteria and 4 category scores.
    Abbr. of 12 {sub-criteria}: CC = Character Consistency, PA = Physical Authenticity,
    LS = Lip Sync, VH = Visual Harmony;
    SC = Shot Continuity, AC = Audio--Visual Correlation;
    MT = Musical Theme, ST = Storytelling, EM = Emotional Expression; 
    VQ = Visual Quality, CR = Creativity, and AN = AI Novelty.
    Abbr. of {categories}: Tech = Technical, Post = Post-Production, and Cont = Content.
    All scores are in $[1,5]$ (higher is better).
  }
  \label{tab:Gemini-2.5-Pro_subscores}
  \centering
  \setlength{\tabcolsep}{3pt}
  \resizebox{\linewidth}{!}{%
  \begin{tabular}{l|cccc|cc|ccc|ccc|cccc}
    \toprule
    Method &
    CC & PA & LS & VH &
    SC & AC &
    MT & ST & EM &
    VQ & CR & AN &
    Tech & Post & Cont & Art \\
    \midrule
    OpenArt-story &
    4.60 & 4.10 & 3.33 & 4.90 &
    4.10 & 4.60 &
    4.70 & 3.47 & 4.10 &
    4.83 & 3.47 & 4.43 &
    4.23 & 4.35 & 4.09 & 4.24 \\
    AutoMV (full) &
    4.07 & 4.30 & 4.37 & 4.47 &
    4.30 & 4.80 &
    4.90 & 4.13 & 4.73 &
    4.70 & 3.27 & 2.87 &
    4.30 & 4.55 & 4.59 & 3.61 \\
    \midrule
    Human (experts) &
    4.93 & 4.63 & 4.67 & 4.83 &
    4.50 & 4.90 &
    5.00 & 3.87 & 4.70 &
    4.70 & 3.90 & 1.00 &
    4.77 & 4.70 & 4.52 & 3.20 \\
    \bottomrule
  \end{tabular}
    }
\end{table*}

\begin{table*}[!tp]
  \caption{
    Detailed Gemini-3-pro-preview evaluation across 12 sub-criteria and 4 category scores.
    Abbr. of 12 {sub-criteria}: CC = Character Consistency, PA = Physical Authenticity,
    LS = Lip Sync, VH = Visual Harmony;
    SC = Shot Continuity, AC = Audio--Visual Correlation;
    MT = Musical Theme, ST = Storytelling, EM = Emotional Expression; 
    VQ = Visual Quality, CR = Creativity, and AN = AI Novelty.
    Abbr. of {categories}: Tech = Technical, Post = Post-Production, and Cont = Content.
    All scores are in $[1,5]$ (higher is better).
  }
  \label{tab:Gemini-3-pro-preview_subscores}
  \centering
  \setlength{\tabcolsep}{3pt}
  \resizebox{\linewidth}{!}{%
  \begin{tabular}{l|cccc|cc|ccc|ccc|cccc}
    \toprule
    Method &
    CC & PA & LS & VH &
    SC & AC &
    MT & ST & EM &
    VQ & CR & AN &
    Tech & Post & Cont & Art \\
    \midrule
    OpenArt-story &
    3.40 & 2.90 & 2.26 & 4.13 &
    3.30 & 3.73 &
    3.90 & 2.60 & 3.00 &
    4.17 & 2.37 & 3.23 &
    3.17 & 3.52 & 3.17 & 3.26 \\
    AutoMV (full) &
    3.47 & 3.10 & 2.47 & 4.20 &
    3.73 & 4.03 &
    4.27 & 3.33 & 3.73 &
    4.23 & 2.76 & 3.43 &
    3.31 & 3.88 & 3.78 & 3.47 \\
    \midrule
    Human (experts) &
    4.87 & 4.73 & 3.93 & 4.80 &
    4.47 & 4.57 &
    4.73 & 3.50 & 4.33 &
    4.60 & 3.40 & 1.40 &
    4.58 & 4.52 & 4.19 & 3.13 \\
    \bottomrule
  \end{tabular}
    }
\end{table*}

\begin{table*}[!tp]
  \caption{
    Detailed Qwen-Omni-3 evaluation across 12 sub-criteria and 4 category scores.
    Abbr. of 12 {sub-criteria}: CC = Character Consistency, PA = Physical Authenticity,
    LS = Lip Sync, VH = Visual Harmony;
    SC = Shot Continuity, AC = Audio--Visual Correlation;
    MT = Musical Theme, ST = Storytelling, EM = Emotional Expression; 
    VQ = Visual Quality, CR = Creativity, and AN = AI Novelty.
    Abbr. of {categories}: Tech = Technical, Post = Post-Production, and Cont = Content.
    All scores are in $[1,5]$ (higher is better).
  }
  \label{tab:Qwen-Omni-3_subscores}
  \centering
  \setlength{\tabcolsep}{3pt}
  \resizebox{\linewidth}{!}{%
  \begin{tabular}{l|cccc|cc|ccc|ccc|cccc}
    \toprule
    Method &
    CC & PA & LS & VH &
    SC & AC &
    MT & ST & EM &
    VQ & CR & AN &
    Tech & Post & Cont & Art \\
    \midrule
    OpenArt-story &
    4.00 & 4.00 & 4.00 & 4.00 &
    3.00 & 3.33 &
    4.00 & 4.00 & 4.00 &
    4.00 & 4.00 & 4.00 &
    4.00 & 3.17 & 4.00 & 4.00 \\
    AutoMV (full) &
    3.93 & 3.93 & 3.93 & 4.00 &
    3.00 & 3.07 &
    3.63 & 4.00 & 4.00 &
    4.00 & 3.97 & 4.00 &
    3.95 & 3.04 & 3.88 & 3.99 \\
    \midrule
    Human (experts) &
    4.00 & 3.80 & 4.00 & 4.00 &
    3.00 & 3.13 &
    3.90 & 4.07 & 4.07 &
    4.07 & 4.00 & 4.00 &
    3.95 & 3.07 & 4.01 & 4.02 \\
    \bottomrule
  \end{tabular}
    }
\end{table*}

\begin{table*}[!tp]
  \caption{
    Detailed Qwen-Omni-2.5 evaluation across 12 sub-criteria and 4 category scores.
    Abbr. of 12 {sub-criteria}: CC = Character Consistency, PA = Physical Authenticity,
    LS = Lip Sync, VH = Visual Harmony;
    SC = Shot Continuity, AC = Audio--Visual Correlation;
    MT = Musical Theme, ST = Storytelling, EM = Emotional Expression; 
    VQ = Visual Quality, CR = Creativity, and AN = AI Novelty.
    Abbr. of {categories}: Tech = Technical, Post = Post-Production, and Cont = Content.
    All scores are in $[1,5]$ (higher is better).
  }
  \label{tab:Qwen-Omni-2.5_subscores}
  \centering
  \setlength{\tabcolsep}{3pt}
  \resizebox{\linewidth}{!}{%
  \begin{tabular}{l|cccc|cc|ccc|ccc|cccc}
    \toprule
    Method &
    CC & PA & LS & VH &
    SC & AC &
    MT & ST & EM &
    VQ & CR & AN &
    Tech & Post & Cont & Art \\
    \midrule
    OpenArt-story &
    4.80 & 4.73 & 4.80 & 4.93 &
    4.57 & 4.70 &
    5.00 & 5.00 & 5.00 &
    5.00 & 4.16 & 4.20 &
    4.82 & 4.64 & 5.00 & 4.45 \\
    AutoMV (full) &
    4.83 & 4.83 & 4.90 & 4.97 &
    4.67 & 4.77 &
    5.00 & 5.00 & 5.00 &
    5.00 & 4.07 & 4.23 &
    4.88 & 4.72 & 5.00 & 4.43 \\
    \midrule
    Human (experts) &
    4.77 & 4.70 & 4.77 & 4.70 &
    4.20 & 4.07 &
    4.87 & 4.97 & 4.87 &
    5.00 & 4.03 & 4.20 &
    4.74 & 4.14 & 4.90 & 4.41 \\
    \bottomrule
  \end{tabular}
    }
\end{table*}

\begin{table*}[!tp]
  \caption{
    Detailed Gemini-3.1-Pro (official API, full-length video at 1\,fps) evaluation across 12 sub-criteria and 4 category scores.
    Abbr. of 12 {sub-criteria}: CC = Character Consistency, PA = Physical Authenticity,
    LS = Lip Sync, VH = Visual Harmony;
    SC = Shot Continuity, AC = Audio--Visual Correlation;
    MT = Musical Theme, ST = Storytelling, EM = Emotional Expression;
    VQ = Visual Quality, CR = Creativity, and AN = AI Novelty.
    Abbr. of {categories}: Tech = Technical, Post = Post-Production, and Cont = Content.
    All scores are in $[1,5]$ (higher is better).
  }
  \label{tab:Gemini-3.1-Pro_subscores}
  \centering
  \setlength{\tabcolsep}{3pt}
  \resizebox{\linewidth}{!}{%
  \begin{tabular}{l|cccc|cc|ccc|ccc|cccc}
    \toprule
    Method &
    CC & PA & LS & VH &
    SC & AC &
    MT & ST & EM &
    VQ & CR & AN &
    Tech & Post & Cont & Art \\
    \midrule
    OpenArt-story &
    3.33 & 2.53 & 2.10 & 4.03 & 3.10 & 3.13 & 3.47 & 2.47 & 2.87 & 3.83 & 2.27 & 2.63 & 3.00 & 3.12 & 2.93 & 2.91 \\
    AutoMV (full) &
    2.33 & 2.53 & 2.20 & 3.57 & 3.03 & 3.53 & 3.83 & 2.93 & 3.20 & 3.63 & 2.30 & 2.60 & 2.66 & 3.28 & 3.32 & 2.84 \\
    \midrule
    Human (experts) &
    4.53 & 4.37 & 4.07 & 4.67 & 4.37 & 4.63 & 4.67 & 3.93 & 4.40 & 4.60 & 3.70 & 1.73 & 4.41 & 4.50 & 4.33 & 3.34 \\
    \bottomrule
  \end{tabular}
    }
\end{table*}

\begin{table*}[!tp]
  \caption{
    Detailed Gemini-3.5-Flash evaluation (official API, full-length video with audio, temperature~0) across 12 sub-criteria and 4 category scores; these are the numeric values behind \cref{fig:judge_radar}.
    Abbr. of 12 {sub-criteria}: CC = Character Consistency, PA = Physical Authenticity,
    LS = Lip Sync, VH = Visual Harmony;
    SC = Shot Continuity, AC = Audio--Visual Correlation;
    MT = Musical Theme, ST = Storytelling, EM = Emotional Expression;
    VQ = Visual Quality, CR = Creativity, and AN = AI Novelty.
    Abbr. of {categories}: Tech = Technical, Post = Post-Production, and Cont = Content.
    All scores are in $[1,5]$ (higher is better).
  }
  \label{tab:Gemini-3.5-Flash_subscores}
  \centering
  \setlength{\tabcolsep}{3pt}
  \resizebox{\linewidth}{!}{%
  \begin{tabular}{l|cccc|cc|ccc|ccc|cccc}
    \toprule
    Method &
    CC & PA & LS & VH &
    SC & AC &
    MT & ST & EM &
    VQ & CR & AN &
    Tech & Post & Cont & Art \\
    \midrule
    Music2Video~\citep{kim2022music2video} &
    1.10 & 1.10 & 1.07 & 3.43 & 2.67 & 3.03 & 2.50 & 1.17 & 2.23 & 2.43 & 2.60 & 3.53 & 1.67 & 2.85 & 1.97 & 2.86 \\
    MV-Crafter~\citep{chen2025mv} &
    2.91 & 3.27 & 1.36 & 3.82 & 3.36 & 4.00 & 4.27 & 2.82 & 3.55 & 4.27 & 3.00 & 3.55 & 2.84 & 3.68 & 3.55 & 3.61 \\
    CHARCHA~\citep{agarwal2025secure} &
    3.24 & 3.20 & 2.20 & 4.16 & 3.60 & 4.32 & 4.20 & 3.04 & 3.52 & 4.20 & 3.76 & 4.00 & 3.20 & 3.96 & 3.59 & 3.99 \\
    Revid.ai-base &
    3.72 & 3.45 & 2.12 & 4.32 & 3.79 & 4.32 & 4.42 & 3.55 & 3.92 & 4.42 & 3.62 & 4.06 & 3.40 & 4.05 & 3.96 & 4.03 \\
    OpenArt-story &
    4.40 & 3.73 & 3.43 & 4.77 & 4.10 & 4.43 & 4.43 & 3.67 & 4.00 & 4.73 & 3.80 & 4.13 & 4.08 & 4.27 & 4.03 & 4.22 \\
    AutoMV (full) &
    4.44 & 4.22 & 4.45 & 4.72 & 4.54 & 4.91 & 4.92 & 4.28 & 4.53 & 4.75 & 4.10 & 3.90 & 4.46 & 4.72 & 4.58 & 4.25 \\
    \midrule
    Human (experts) &
    4.93 & 4.90 & 4.90 & 4.97 & 4.97 & 5.00 & 4.97 & 4.47 & 4.90 & 4.97 & 4.57 & 1.20 & 4.93 & 4.98 & 4.78 & 3.58 \\
    \bottomrule
  \end{tabular}
    }
\end{table*}

\begin{table*}[!tp]
  \caption{
    Detailed Qwen3.5-Omni evaluation across 12 sub-criteria and 4 category scores.
    Unlike Qwen-Omni-2.5/3, Qwen3.5-Omni is fed the full-length (compressed to 480p/8fps with audio retained) video rather than a 30-second clip.
    Abbr. of 12 {sub-criteria}: CC = Character Consistency, PA = Physical Authenticity,
    LS = Lip Sync, VH = Visual Harmony;
    SC = Shot Continuity, AC = Audio--Visual Correlation;
    MT = Musical Theme, ST = Storytelling, EM = Emotional Expression;
    VQ = Visual Quality, CR = Creativity, and AN = AI Novelty.
    Abbr. of {categories}: Tech = Technical, Post = Post-Production, and Cont = Content.
    All scores are in $[1,5]$ (higher is better).
  }
  \label{tab:Qwen3.5-Omni_subscores}
  \centering
  \setlength{\tabcolsep}{3pt}
  \resizebox{\linewidth}{!}{%
  \begin{tabular}{l|cccc|cc|ccc|ccc|cccc}
    \toprule
    Method &
    CC & PA & LS & VH &
    SC & AC &
    MT & ST & EM &
    VQ & CR & AN &
    Tech & Post & Cont & Art \\
    \midrule
    OpenArt-story &
    3.10 & 3.38 & 1.38 & 4.48 & 3.66 & 3.45 & 4.17 & 3.07 & 3.86 & 4.59 & 3.55 & 4.45 & 3.09 & 3.55 & 3.70 & 4.20 \\
    AutoMV (full) &
    4.07 & 3.38 & 4.48 & 4.66 & 4.21 & 4.69 & 4.69 & 4.14 & 4.62 & 4.69 & 4.07 & 4.52 & 4.15 & 4.45 & 4.48 & 4.43 \\
    \midrule
    Human (experts) &
    4.17 & 3.83 & 4.37 & 4.77 & 4.30 & 4.70 & 4.80 & 4.13 & 4.63 & 4.57 & 4.23 & 3.43 & 4.28 & 4.50 & 4.52 & 4.08 \\
    \bottomrule
  \end{tabular}
    }
\end{table*}

{The judge configurations differ in score range, saturation, and category-level ordering.}

\textbf{Gemini-2.5-Flash and 2.5-Pro.}
{\cref{tab:Gemini-2.5-Flash_subscores,tab:Gemini-2.5-Pro_subscores} show that both judges rank human-directed MVs first, AutoMV second, and OpenArt third in Technical and Post-Production. In Content, AutoMV approaches or slightly exceeds the human-directed means. OpenArt receives higher AI Novelty scores in some configurations, while AutoMV scores closer to human-directed MVs on VQ and EM. Gemini-2.5-Pro generally assigns lower scores than Flash.}

\textbf{Gemini-3-Pro-Preview.}
{\cref{tab:Gemini-3-pro-preview_subscores} shows lower absolute scores and a narrower range than Gemini-2.5-Pro. Human-directed MVs receive the highest category means, and AutoMV scores above OpenArt in Content and Post-Production.}

\textbf{Qwen-Omni-3 (30-second clips).}
{In \cref{tab:Qwen-Omni-3_subscores}, OpenArt, AutoMV, and human-directed MVs receive similar scores, often differing by 0.1--0.3 points. Many criterion means lie between 3.8 and 4.1 under the 30-second input protocol.}

\textbf{Qwen-Omni-2.5 (30-second clips).}
{Table~\ref{tab:Qwen-Omni-2.5_subscores} shows stronger saturation: most criterion means lie between 4.7 and 5.0, with the aggregate ordering Human $\approx$ AutoMV $\geq$ OpenArt.}

\textbf{Gemini-3.1-Pro and Gemini-3.5-Flash.}
{\cref{tab:Gemini-3.1-Pro_subscores,tab:Gemini-3.5-Flash_subscores} report full-video scores. Gemini-3.1-Pro assigns AutoMV lower Technical scores than OpenArt ($2.66$ vs.\ $3.00$), and higher Post-Production ($3.28$ vs.\ $3.12$) and Content scores ($3.32$ vs.\ $2.93$). AutoMV leads on audio- and narrative-related criteria, while OpenArt scores higher on CC, VH, and VQ. Human-directed MVs receive an AI Novelty score of $1.73$.}

{Gemini-3.5-Flash, our primary judge, covers all seven systems in \cref{tab:Gemini-3.5-Flash_subscores} and \cref{fig:judge_radar}. AutoMV leads the automatic systems in Technical, Post-Production, and Content, scoring $4.58$ on Content against $4.78$ for human-directed MVs. OpenArt receives the highest AI Novelty score ($4.13$). Both Gemini judges assign human-directed MVs the lowest AI Novelty score.}

\textbf{Qwen3.5-Omni (full-length video).}
{Qwen3.5-Omni (\cref{tab:Qwen3.5-Omni_subscores}) processes the full-length compressed video and ranks AutoMV above OpenArt on Technical, Post-Production, and Content. It assigns human-directed MVs the lowest AI Novelty score and generally assigns higher absolute scores than Gemini-3.1-Pro.}

{Song-grouped human agreement and judge selection across these configurations are reported in \cref{app:llm_human_relation}.}

\section{Failure Case Study}
{We analyze four recurring failure modes in AutoMV outputs: physical implausibility, off-beat dance movements, textual inconsistencies, and lip-sync mismatch.}
\begin{figure}[ht]
    \centering
    \includegraphics[width=0.7\linewidth]{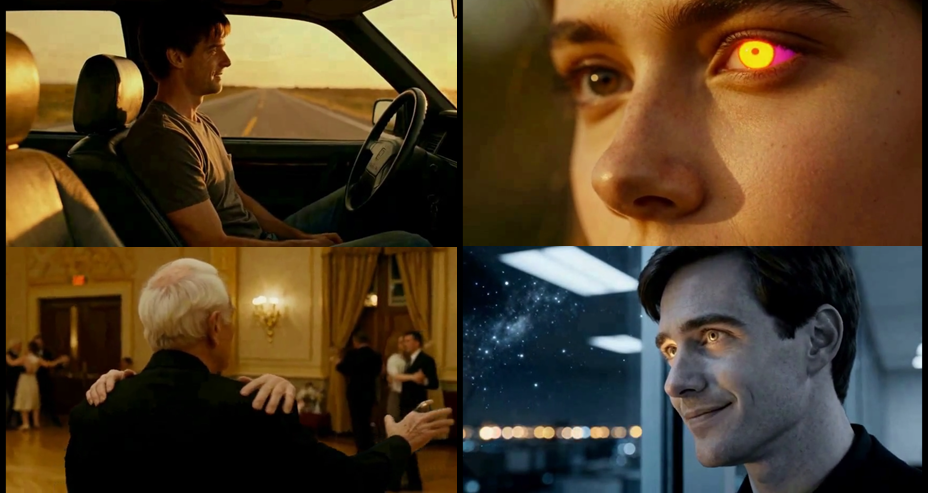}
    \caption{Examples of physically implausible generations}
    \label{fig:unrealistic}
\end{figure}

\textbf{Physically Implausible}. The pipeline occasionally generates unrealistic outputs, including abnormal hand-object contacts, glowing eyes, and anatomically implausible human poses.

\textbf{Off-Beat Dance Movements}. Dance motion is not explicitly synchronized to musical beats in the current pipeline. Beat-tracking and rhythm-alignment modules are not included.
\begin{figure}[ht]
    \centering
    \includegraphics[width=0.7\linewidth]{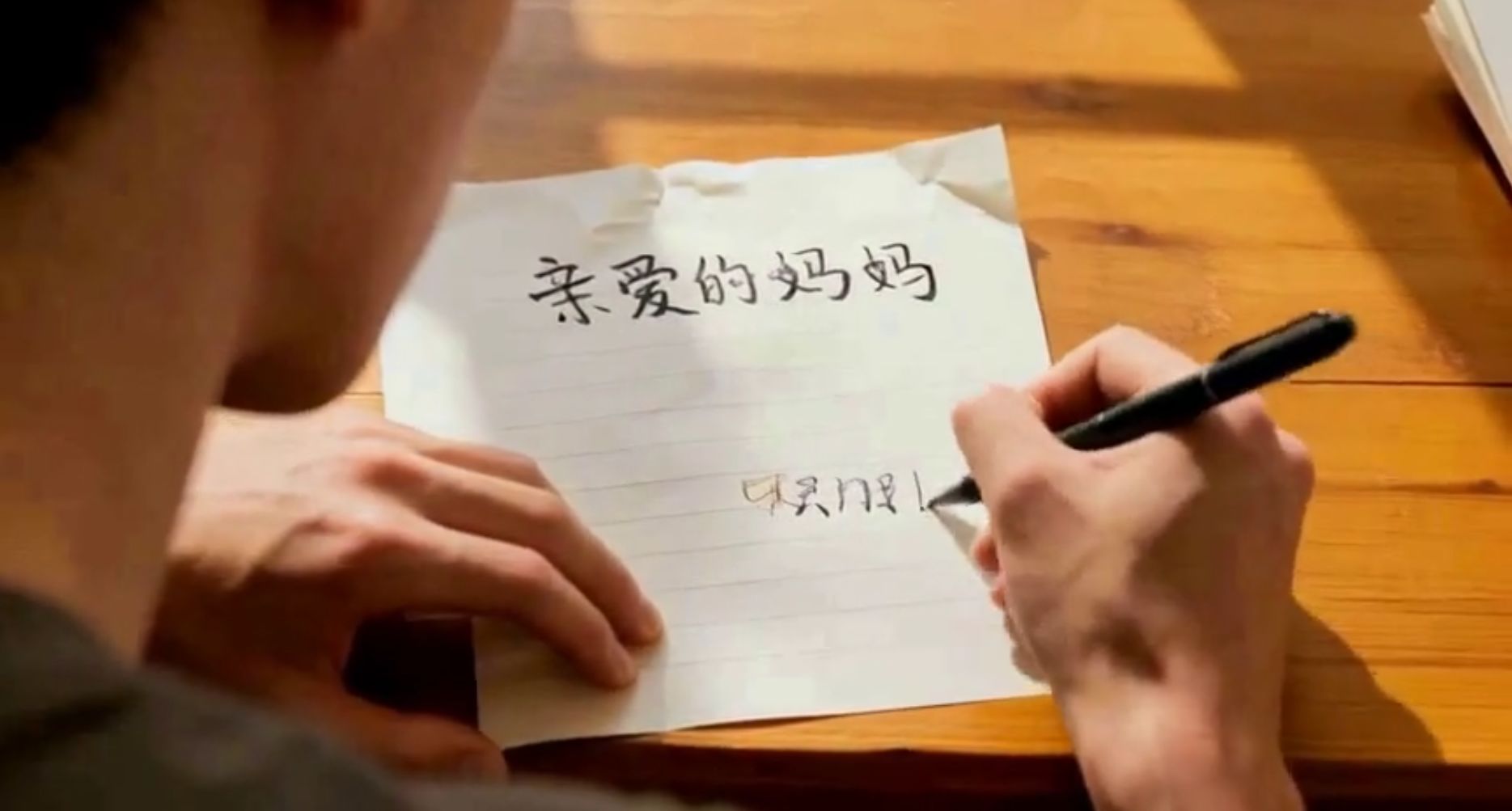} 
    \caption{Handwritten letter close-up with inconsistent glyph shapes (the second line in the image) and temporal progression.}
    \label{fig:text_error}
\end{figure}

\textbf{Textual Inconsistencies}. On-screen text can contain inconsistent glyph shapes and temporal progression, especially in handwritten-letter shots. Whisper transcription errors can propagate to downstream text and subtitle generation, producing distorted, illegible, or audio-misaligned text.
\begin{figure}[ht]
    \centering
    \includegraphics[width=0.7\linewidth]{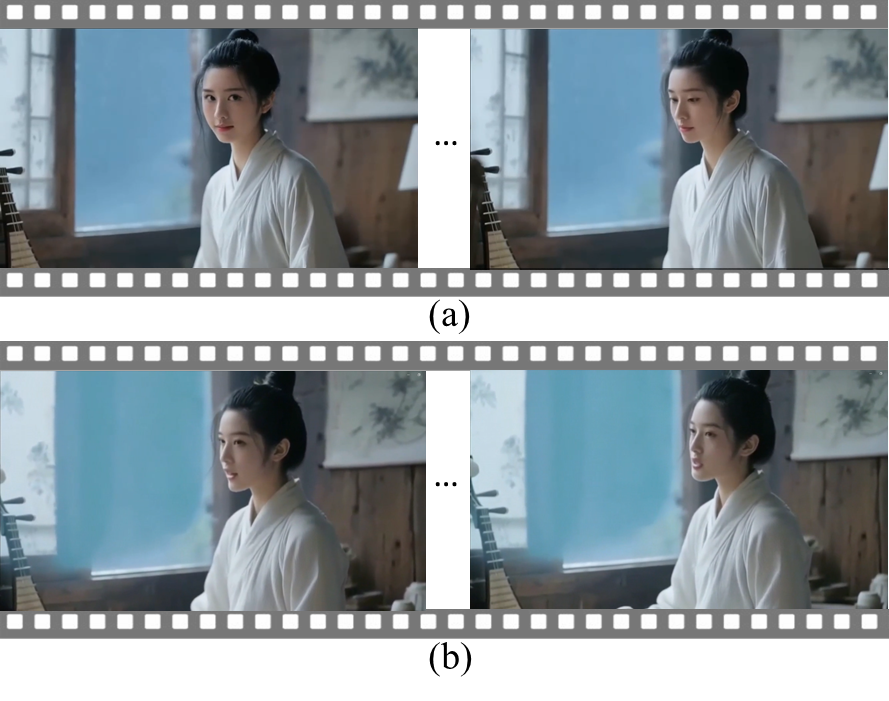}
    \caption{(a) without source separation(b) source separation
}
    \label{fig:lip}
\end{figure}

\textbf{Lip-Sync mismatch without source separation}. The lip-sync model processes mixed audio tracks, causing visible misalignment between mouth movements and lyrical syllables, which is particularly pronounced in Jay Chou's songs as shown in \cref{fig:lip}. Lip-syncing with source separation significantly outperforms the non-separation approach in both accuracy and naturalness.Thus, source separation is employed in our main experiments.
\begin{table*}[!tp]
\centering
\small
\caption{\textbf{Quantitative comparison of different baseline methods on the VBench benchmark.} We evaluate across five core video generation dimensions: subject consistency, background consistency, aesthetic quality, imaging quality, and human anatomy. }
\label{tab:vbench_comparison}
\begin{tabular}{lccccccccc}
\toprule
\textbf{Metric} & \textbf{GT} & \textbf{AutoMV} & \textbf{OpenArt} & \textbf{Revid} & \textbf{music2video} & \textbf{mv-crafter} & \textbf{avlink} & \textbf{charcha} & \textbf{yingvideo} \\
\midrule
Subject Consistency & 0.8353 & \textbf{0.9211} & 0.9033 & 0.9080 & 0.8873 & 0.9058 & 0.9313 & 0.8294 & 0.9443 \\
Background Consistency & 0.9247 & \textbf{0.9382} & 0.9380 & 0.9370 & 0.9054 & 0.9455 & 0.9518 & 0.8963 & 0.9448 \\
Aesthetic Quality & 0.5212 & 0.5678 & 0.5765 & 0.5861 & 0.3876 & \textbf{0.6751} & 0.5008 & 0.6015 & 0.5417 \\
Imaging Quality & 0.4836 & 0.6143 & \textbf{0.6535} & 0.6275 & 0.6242 & 0.6128 & 0.6705 & 0.6454 & 0.6237 \\
Human Anatomy & 0.9251 & \textbf{0.9496} & 0.9469 & 0.8681 & \textbf{0.6903} & 0.7968 & 0.8786 & 0.6880 & 0.9511 \\
\bottomrule
\end{tabular}
\end{table*}
\section[VBench Metric Behavior on Music Videos]{VBench Metric Behavior on Music Videos}\label{app:vbench}

{Several automatic systems score above the human-directed reference on VBench~\cite{huang2024vbench} metrics (\cref{tab:vbench_comparison}), while expert overall scores favor the human-directed MVs (\cref{tab:human_subscores}). We examine how the visual properties measured by these metrics relate to music-video editing and motion.}

{\textbf{Subject and Background Consistency.} These metrics compare visual features across frames. The human-directed reference scores $0.8353$ on Subject Consistency and $0.9247$ on Background Consistency. Shot cuts, camera movement, and changing poses can lower feature similarity, while static subjects and backgrounds preserve it. This sensitivity connects consistency scores to editing and motion as well as identity preservation.}

{\textbf{Aesthetic and Imaging Quality.} The human-directed reference scores $0.5212$ and $0.4836$ on these image-level predictors. Music videos use low light, motion blur, film grain, and stage lighting as visual devices; the predictors' response to such styles offers a possible explanation for their different ordering from expert overall ratings.}

{\textbf{Human Anatomy.} Anatomy assessment depends on visible body structure. Dance, unusual poses, partial occlusion, and costumes change that visibility. Inspecting these cases can help distinguish body-generation errors from detector sensitivity to performance footage.}

\section{Variability and Significance Testing}\label{app:significance}

{AutoMV, OpenArt, Revid, and the human-directed reference each contribute 30 outputs. \cref{tab:significance} reports per-song means and standard deviations and Holm-adjusted paired Wilcoxon comparisons against the commercial baselines. Pairing by song accounts for the common musical input.}

\begin{table*}[!tp]
  \caption{Mean\,$\pm$\,standard deviation over the 30 benchmark songs. Against the two commercial baselines we report Holm-adjusted two-sided paired Wilcoxon $p$-values for comparisons significant at $p<0.05$ (paired by song, $n{=}30$). ``Judge'' is the weighted Gemini-3.5-Flash score on the $[1,5]$ rubric. $^{*}p<0.05$, $^{**}p<0.01$, $^{***}p<0.001$.}
  \label{tab:significance}
  \centering
  \small
  \begin{tabular}{l|cccc|cc}
    \toprule
    & \multicolumn{4}{c|}{Mean $\pm$ std over the 30 songs} & \multicolumn{2}{c}{AutoMV vs.\ commercial baselines} \\
    Metric & Human & AutoMV & OpenArt & Revid & OpenArt & Revid \\
    \midrule
    IB\,(\%) & $24.9\pm6.0$ & $24.9\pm6.4$ & $18.8\pm6.6$ & $20.3\pm6.9$ & $2.1\!\times\!10^{-4\,***}$ & $1.5\!\times\!10^{-3\,**}$ \\
    Beat\,(\%) & $26.1\pm5.4$ & $26.7\pm5.3$ & $24.1\pm5.2$ & $17.0\pm5.1$ & $0.038^{*}$ & $7.4\!\times\!10^{-7\,***}$ \\
    Judge & $4.5\pm0.1$ & $4.5\pm0.3$ & $4.1\pm0.3$ & $3.9\pm0.5$ & $2.1\!\times\!10^{-3\,**}$ & $1.9\!\times\!10^{-5\,***}$ \\
    \bottomrule
  \end{tabular}
\end{table*}

{AutoMV scores higher than both commercial baselines on IB, Beat, and the LLM-judge score at Holm-adjusted $p<0.05$, with matched-pairs rank-biserial effect sizes of $0.57$--$0.96$. The standard deviations show substantial variation across songs.}

{Gemini-3.5-Flash awards the maximum rating to $97$--$100\%$ of human-directed MVs on five of the twelve criteria; their weighted scores have a per-song standard deviation of $0.14$. Expert annotations give weighted means of $\ExpertGroundTruthWeightedMean$ for human-directed MVs and $\ExpertAutoMVWeightedMean$ for AutoMV (\cref{tab:human_subscores}).}

\section[Additional Audio-Visual Metric Diagnostics]{Additional Audio-Visual Metric Diagnostics}\label{app:failed_metrics}

{We evaluated nine additional audio-visual metrics on the same benchmark (\cref{tab:failed_metrics}). Expert means rank human-directed MVs first ($\ExpertGroundTruthWeightedMean$, followed by AutoMV $\ExpertAutoMVWeightedMean$, OpenArt $\ExpertOpenArtWeightedMean$, and Revid $\ExpertRevidWeightedMean$). Eight of the nine metrics place at least one automatic system above the human-directed reference. CLIP lyric--frame matching ranks the reference first with a max--min range of $0.013$. These aggregate patterns show which properties each metric captures.}

\begin{table*}[!tp]
  \caption{Additional objective metrics, with per-method means on the 30-song benchmark (30 videos per method). Higher is better for all rows; \textbf{bold} marks the top-scoring method. ``GT rank'' is the position of the human-directed MVs among the four methods; their expert overall mean ranks first. ``Spread'' is the max--min gap across method means.}
  \label{tab:failed_metrics}
  \centering
  \small
  \begin{tabular}{l|cccc|cc}
    \toprule
    Metric & GT & AutoMV & OpenArt & Revid & GT rank & Spread \\
    \midrule
    Emotion match (CLIP zero-shot) & 0.33 & 0.43 & \textbf{0.51} & 0.32 & 3rd & 0.190 \\
    Emotion sync (MLLM valence--arousal) & 0.78 & 0.81 & \textbf{0.84} & 0.83 & 4th & 0.060 \\
    Lyric--frame match (CLIP, per line) & \textbf{0.222} & 0.218 & 0.209 & 0.220 & 1st & 0.013 \\
    Lyric--frame match (MLLM judge, 1--5) & 2.87 & \textbf{3.08} & 2.61 & \textbf{3.08} & 3rd & 0.470 \\
    Music--video match (MuLan) & 0.085 & 0.088 & \textbf{0.090} & 0.088 & 4th & 0.005 \\
    Music--emotion-word match (CLAP) & 0.285 & 0.271 & 0.293 & \textbf{0.298} & 3rd & 0.027 \\
    Audio--caption reward model (CMI-RM) & 1.30 & 1.29 & 1.24 & \textbf{1.32} & 2nd & 0.080 \\
    Global onset--motion correlation & 0.033 & 0.013 & $-0.007$ & \textbf{0.061} & 2nd & 0.068 \\
    Cut-on-beat ratio & 0.573 & 0.550 & \textbf{0.599} & 0.582 & 3rd & 0.049 \\
    \bottomrule
  \end{tabular}
\end{table*}

\paragraph{Metric descriptions.}
The \emph{emotion} family measures whether the video's emotional content matches the music: via CLIP zero-shot classification of frames against emotion words compared to the lyrics' emotion (row~1), or via an MLLM rating per-segment valence--arousal trajectories for the visuals and the music and correlating the two (row~2). The \emph{lyric--frame} family measures literal text--visual correspondence per lyric line, either by CLIP embedding similarity between each sung line and its concurrent frames (row~3) or by an MLLM judging the semantic match of each line--shot pair on a 1--5 scale (row~4). The \emph{music--video embedding} family scores the similarity between the song audio and a textual description of the video in a joint audio--text space, using MuLan (row~5), CLAP with single emotion-word text (row~6), or a music--text reward model (row~7). Finally, two \emph{rhythm} variants differ from the validated Beat metric: a global zero-lag correlation between audio onsets and optical flow (row~8), and the fraction of shot cuts landing on beats (row~9).

{\paragraph{Patterns in metric behavior.}
The aggregate scores suggest three factors for closer analysis.}

{\textbf{(i) Literal lyric illustration.} Revid's image slideshows illustrate lyrics with static pictures. Revid ties for first on the MLLM lyric--frame score and ranks first on CLAP emotion-word matching and the audio--caption reward model, while receiving the lowest expert overall mean ($\ExpertRevidWeightedMean$). Literal illustration can raise these correspondence scores even when experts give low ratings for narrative and visual quality. Slide transitions also produce large optical-flow changes, offering a possible explanation for Revid's high global onset--motion correlation.}

{\textbf{(ii) Compressed score ranges.} Emotion sync spans $0.06$ across the four method means, and MuLan similarity spans $0.005$. These metrics assign similar aggregate scores to systems with substantially different expert overall ratings.}

{\textbf{(iii) Different temporal cues.} Cut-on-beat ratio ranks OpenArt first ($0.599$), followed by Revid ($0.582$), human-directed MVs ($0.573$), and AutoMV ($0.550$). It measures the coincidence of cuts and beats; the main-paper Beat metric measures beat-to-motion synchronization within lyric segments. The two metrics therefore track different aspects of temporal alignment.}

{\paragraph{Implementation diagnostics.}
Mood terms have average pairwise embedding cosine similarity of approximately $0.92$; one CLAP checkpoint compresses audio--text similarities into a $\pm0.01$ range; and the MLLM lyric judge uses effectively three values on its 1--5 scale. The reward-model comparison uses a different setting from its training data, and shot-cut detection is noisy on generated videos. Static-matching scores remain weakly aligned with expert ratings, while a time-aligned emotion-sync variant shows a weak positive association.}

{\paragraph{Implications for evaluation.}
The observed differences between local correspondence scores and expert overall ratings motivate evaluating music-video quality across multiple dimensions. M2VBench combines objective measures with a rubric for narrative, musical, visual, and artistic qualities, and uses same-song expert preferences to measure agreement.}

\begingroup
\section{Human Alignment and Grouped Judge Selection}\label{app:llm_human_relation}

We analyze song-level agreement for the weighted-total endpoint and evaluate judge selection with songs held out as groups.

\paragraph{Annotation mapping and common support.}
Videos were presented to experts under shuffled identifiers.
The shuffle files provide an exact mapping between the presented video IDs and the corresponding songs and systems for the 90 GroundTruth, OpenArt, and AutoMV videos from 30 songs. Six full-video judge configurations are compared: Gemini-3.5-Flash, Gemini-2.5-Pro, Gemini-2.5-Flash, Gemini-3.1-Pro at temperatures 0 and 1, and Qwen3.5-Omni. Other Gemini runs use temperature 0. The primary comparison uses the complete 30-song set (90 videos); Qwen runs using short clips and excluding OpenArt are reported separately in \cref{app:llm_score}.

These video-level ratings supply the same-song judge comparisons below. The study's aggregate expert summaries and ablation scores are reported separately in \cref{tab:human_subscores}.

\paragraph{Song-level preference agreement.}
For each song and criterion, we compare two methods' scores and check whether the judge agrees with the experts' strict preference. Human ties are excluded; judge ties count as disagreement. The primary rubric endpoint averages accuracy over the human-decided criteria within each song, then weights the 30 songs equally. Its automatic-method comparison is AutoMV versus OpenArt. A supplementary comparison includes all three methods. The weighted-total endpoint first applies the manuscript's category weights (0.05 per Technical criterion and 0.10 per remaining criterion), then compares the two totals. Ties use a numerical tolerance of $10^{-9}$.

\paragraph{Grouped selection and held-out evaluation.}
This analysis uses responses generated independently of the human annotations, with the models, prompts, rubric weights, and decision thresholds fixed. We shuffle the sorted song IDs with NumPy PCG64 seed 20261009 and use a five-fold split of six songs as the primary analysis. We additionally repeat the same grouped procedure with three folds of ten songs. All videos, criteria, and comparisons from a song remain in the same fold. For each outer fold and endpoint, we select the configuration with the highest training-song agreement and evaluate it on the held-out songs. Ties between configurations use a fixed lexicographic identifier order.

We pool the resulting out-of-fold (OOF) song scores so that every song contributes once. Percentile 95\% intervals use 20,000 bootstrap resamples of whole songs (seed 20261010), preserving dependent criteria and comparisons. We additionally repeat the complete selection procedure for seeds 0--19 on the same 30 songs. Fixed-configuration rows summarize all 30 songs; fold-selected rows summarize held-out performance after selection on the other songs.

\begin{table*}[!tp]
\centering
\caption{Song-grouped held-out agreement for the weighted-total endpoint on 30 complete songs (90 mapped videos). Gemini-3.5-Flash is selected in every fold of both grouped splits. Human ties are excluded and judge ties count as disagreement; intervals are 95\% song-bootstrap intervals.}
\label{tab:judge_groupcv}
\small
\begin{tabular}{lcc}
\toprule
Selection procedure & Held-out agreement (\%) & Selected configuration \\
\midrule
Five-fold grouped selection & 73.3 [55.6, 85.8] & Gemini-3.5-Flash (5/5) \\
Three-fold grouped selection & 73.3 [55.6, 85.8] & Gemini-3.5-Flash (3/3) \\
\bottomrule
\end{tabular}
\end{table*}

\paragraph{Aggregate-score selection.}
For the weighted-total endpoint, Gemini-3.5-Flash is selected in all five primary folds and agrees with the experts on 22 of the 30 held-out AutoMV--OpenArt comparisons: 73.3\% (95\% CI 55.6--85.8\%). The supplementary three-fold split also selects it in all three folds and yields 73.3\% pooled agreement; its fold means are 90.0\%, 60.0\%, and 70.0\%. Across 20 additional seeds, it is selected in all 100 folds from five-fold splits and all 60 folds from three-fold splits, yielding the same pooled agreement. Aggregate-score selection thus returns the same configuration across all evaluated partitions of the mapped AutoMV--OpenArt comparison.

\par
\endgroup

\newpage %
\section{Case Study on Music-Aware Shot Transitions.}
{We manually categorize the primary cue associated with each of 41 shot transitions in one generated full-length video (\cref{tab:transition_cues}). Ten transitions (24.4\%) coincide with lyric endings, eight (19.5\%) with musical section changes, eight (19.5\%) with beats, and three (7.3\%) with downbeats. The remaining 12 (29.3\%) align with other contextual or visual factors. Rhythm-related cues account for 26.8\% of transitions, alongside lyric and section boundaries.}

\section{Use of LLMs}

Large language models (LLMs) were used exclusively for language-related assistance, including editing, formatting, and improving the clarity of the manuscript; (1) all scientific ideas, experimental designs, implementations, and analyses were independently conceived and carried out by the authors; (2) no LLM-generated content was used as experimental data, training data, or reported results in this work.

\begin{table}[!htbp]
    \centering
    \caption{Distribution of the primary cues associated with 41 shot transitions in a generated full-length music video. Musical section changes refer to transitions between structural components of a song, such as verses and choruses. Rhythm-related cues, including beats and downbeats, account for 26.8\% of all transitions.}
    \label{tab:transition_cues}
    \begin{tabular}{lcc}
        \toprule
        \textbf{Transition Cue} & \textbf{Count} & \textbf{Percentage} \\
        \midrule
        
        Lyric ending                    & 10 & 24.4\% \\
        Musical section change          & 8  & 19.5\% \\
        Beat                            & 8  & 19.5\% \\
        Downbeat                        & 3  & 7.3\% \\
        Other factors                   & 12 & 29.3\% \\
        \midrule
        Total                           & 41 & 100.0\% \\
        \bottomrule
    \end{tabular}
\end{table}

\end{document}